\documentclass[aps,floats,showpacs,amssymb,tightenlines]{revtex4}

\usepackage{amsmath}
\usepackage{amsfonts}
\usepackage{amscd}
\usepackage{epsfig}
\usepackage{amssymb}
\usepackage{tabularx}
\usepackage{longtable}
\usepackage{calligra}
\def\a{\alpha}
\def\b{\beta}

\def\e{\epsilon}
\def\I{{\cal I}}

\def\d{\delta}

\def\w{\omega}

\def\p{\partial}

\def\I{{\cal I}}

\def\ri{r_{i}}
\def\ro{r_{o}}
\def\rn{Reissner-Nordstr\"om }

\begin{document}
\title{Gravitational instability  of the inner static region of a Reissner-Nordstr\"om black hole}

\author{Gustavo Dotti and Reinaldo J. Gleiser}
\affiliation {Facultad de Matem\'atica, Astronom\'{\i}a y
F\'{\i}sica (FaMAF), Universidad Nacional de C\'ordoba and\\
Instituto de F\'{\i}sica Enrique Gaviola, CONICET.\\ Ciudad
Universitaria, (5000) C\'ordoba,\ Argentina}

\begin{abstract}
Reissner--Nordstr\"om black holes have two static regions:
 $r > \ro$ and $0 < r < \ri$, where $\ri$ and  $\ro$ are
 the inner and outer horizon radii. The stability of the exterior
 static region has been established long time ago. In this work we prove that the interior
static  region
 is unstable under linear gravitational perturbations, by showing 
that  field perturbations compactly supported within this region
will generically excite a mode that grows exponentially in time.
This result gives an alternative reason to mass inflation
to consider the space time extension beyond the Cauchy horizon as physically irrelevant,
and thus provides support to the strong cosmic censorship conjecture, which is also 
backed  by recent evidence of a  linear gravitational instability
 in the interior region of Kerr black holes found by the authors.
The use of intertwiners to solve for the evolution of initial data plays a key role, 
and adapts without change to the case of super-extremal \rn black holes, allowing to complete
the proof of the linear instability of this naked singularity. A particular intertwiner
is found such that the intertwined Zerilli field has a geometrical meaning -it is  the first order 
variation of a particular Riemann tensor invariant-. 
Using this, calculations can be carried out 
explicitely for every harmonic number. 

 \end{abstract}


\maketitle


\section{Introduction}

In the course of a program \cite{dottigleiser,doglepu,dgsv,nm} to
 study the stability under linear gravitational perturbations of the most notable nakedly singular
 solutions of Einstein's field equations, namely, negative mass Schwarzschild's solution
 \cite{dottigleiser,ghi,nm},
 $|Q| > M > 0$  Reissner--Nordstr\"om space-time \cite{doglepu} and $|J| > M^2$
Kerr space-time \cite{doglepu,dgsv} (see also \cite{cardoso}), we noticed that the stationary interior region
beyond the inner horizon of a Kerr {\em black hole}  is unstable \cite{dgsv,proc}. The existence of an
 initially bounded and exponentially
growing solution of Teukolsky equations in the ``super-extreme" case $|J| > M^2$,
of which some numerical evidence had been given earlier    in
\cite{doglepu}, was  established in \cite{dgsv}, where it was  shown that
there are actually infinitely many unstable modes. It was also shown
in \cite{dgsv}, that the stationary  region beyond the inner horizon
of a  Kerr {\em black hole} (i.e. $|J| \leq M^2$) is linearly unstable under gravitational perturbations.
These results show that linear perturbation theory is a valuable tool  to study not only  weak cosmic
censorship  (impossibility of formation of naked singularities), but also
 strong cosmic
censorship  (impossibility of formation of Cauchy horizons).

For the Kerr spacetime, an explicit expression for the unstable modes is not given in \cite{dgsv},
since they involve solutions of  complicated second order
ordinary differential equations, which can at best  be
 written in terms of Heun functions, providing little extra information.
This, added to the complexity of the reconstruction of the perturbed metric  from a solution of Teukolski's equations,
makes it extremely difficult to evaluate the physical meaningfulness of these unstable modes.\\

The situation is different for the  negative mass Schwarzschild and
the  super-extremal Reissner--Nordstr\"om space-times, where
 explicit expressions for some unstable modes, which involve only elementary functions,  are given in
 \cite{dottigleiser,nm} and \cite{doglepu}  respectively. Since  the metric reconstruction process in the spherically symmetric
case is much simpler,  it is possible to study the effect of perturbations on the singularity, by calculating the perturbed Riemann tensor invariants.
We use these to select appropriate boundary conditions at the singularity that ensure the
self-consistency of the linear perturbation scheme, by requiring
that curvature scalars do not get corrections that diverge faster at the singularity than the zeroth order term.
As an example, in the case of the Schwarzschild negative mass naked singularity there are infinitely many
possible boundary conditions at the singularity, parameterized by $S^1$ \cite{ghi,dottigleiser}, only one of
which satisfies the above requirement. Thus, besides assuring the self-consistency of the perturbative treatment,
the above procedure solves the problem of having a unique, well defined evolution of perturbations
in a non globally hyperbolic background.\\

  The unstable modes in \cite{dottigleiser}
 were
 recognized by Cardoso and Cavaglia  \cite{cardoso} to correspond to Chandrasekar's
  ``algebraic special" (AS)
 solutions of the linearized Einstein's equations \cite{prs,chandra}. This observation
 hinted in the right direction  where to look for unstable modes of
 super-extremal Reissner-Nordstr\"om black
 holes \cite{cardoso,doglepu}.
  As shown in \cite{doglepu}, some of the Reissner--Nordstr\"om AS modes grow
 exponentially in time  while keeping appropriate spatial boundary
 conditions in the super-extremal case, as happens in the negative mass Schwarzschild case.
  With no exception,  the AS solutions are irrelevant
 to the stability problem of the {\em exterior region of
  black holes}, since they do not satisfy suitable
 boundary conditions, a  probable reason why they remained unnoticed for such a long time.
For the Kerr solution, the AS modes do not satisfy appropriate boundary conditions, neither for the black hole
stationary regions, nor
for the naked singularity. However, it was proved in  \cite{dgsv} that unstable modes
exist for every harmonic (i.e., spin weighted spheroidal harmonic) of the Teukolski equations
in the super-extreme case.
Moreover, in \cite{dgsv} a connection was spot
between the unstable modes of Kerr naked singularities,
and unstable modes for the interior stationary region ($r < \ri$) of a Kerr {\em black hole}.
Given the similarities in the structures of the maximal analytic extensions of
Kerr and Reissner-Nordstr\"om black holes, and the fact that both are affected by  Cauchy horizon  issues,
one is  naturally led to ask whether the interior region of a
Reissner-Nordstr\"om black hole is also unstable.
In this paper we  show
 that this is the case.
 We  give a detailed proof of the instability of the inner region of a
\rn black hole  under linear gravitational perturbations initially restricted to a compact subset
of the inner region. \\

This provides  an alternative reason to the mass inflation mechanism to disregard
the extension of the space time manifold beyond the Cauchy horizon: since the inner region is static but unstable,
it cannot be the endpoint of an evolving space time. A similar result for the Kerr black hole
would imply cutting off the inner  region of this space time, which  has closed time like curves and
other pathologies.\\

We  will concentrate on type one  (also called
``gravitational'', as opposed to type two or ``electromagnetic'') scalar (also called ``polar'')
linear perturbations of the metric and electromagnetic fields around the \rn solution, since it is in this sector that
we have found explicit unstable modes.
The linearized Einstein's equations for these type of perturbations can be reduced to a $1+1$ wave equation
on a field $\Phi_1^+$ in a semi-infinite domain bounded by the singularity world-line, with a time independent
potential (Zerilli's equation \cite{reggewheeler,moncrief,kodamaishibashi}). This formalism was used
to prove the stability of the {\em exterior static region} \cite{moncrief} of the \rn black hole. In this case,
Zerilli's equation can be written as  a wave equation in a complete  $1+1$ Minkowski spacetime,
 with a nonsingular
potential which, being positive definite, guarantees the stability under
this kind of gravitational perturbations \cite{moncrief}. \\
When applied to
the static black hole inner region $r < \ri$, instead,
one gets a wave equation on a half $1+1$ fiducial Minkowski spacetime bounded by the singularity worldline,
and the potential
has an unexpected second order pole at an inner  point in the domain (that we call ``kinematic singularity''),
besides the expected  divergence at the  singularity. This makes
the initial value problem for the inner region far more difficult than that for  $r>\ro$.
These technical difficulties, however,
are entirely analogous to those that arise  in the linear perturbation problem of a negative mass
Schwarzschild spacetime, a problem which was solved recently in \cite{nm}. As in the Schwarzschild case, the second order
pole in the potential can be traced back  to the fact that the Zerilli field $\Phi_1^+$, as defined, is a singular function
of the perturbed metric and electromagnetic fields at the kinematic singularity (from where the name ``kinematic" comes).
Thus, an alternative field $\hat \Phi$ has
to be introduced  to properly analyze perturbations \cite{nm}.
This is related to $\Phi_1^+$  by an intertwiner operator: $\hat \Phi = \I \Phi_1^+$, where  $\I = \p/\p x + g$,
and $x$ is a tortoise radial coordinate.
In terms of $\hat \Phi$, the type one scalar gravitational perturbation equation
is a $1+1$ wave equation $\p^2 \hat \Phi/\p t ^ 2 + \hat {\cal H} \hat \Phi=0$, $\hat {\cal H} =
  \p^ 2 /\p {x}^2 + \hat V(x)$,
with a potential $\hat V$ that is regular  everywhere. Once an appropriate self-adjoint extension
of $\hat {\cal H}$ is chosen -and, as explained above, there is a unique physically motivated choice-,
the evolution of initial data  is unambiguously defined by means of an $\hat {\cal H}$ mode expansion of the data.
This gives a dynamics in spite of
the fact that the background is non globally hyperbolic (see \cite{wi} for a similar approach).
 The intertwining technique and
choice of self-adjoint extension is explained  in detail in Section \ref{sinter}. Previously, in  Section  \ref{lp},
we review of the basics of the Reissner-Nordstr\"om solution, its linear perturbations,
the factorization of Zerilli's Hamiltonian, and Chandrasekhar's algebraic special modes.\\

Several technical aspects of the problem are dealt with in the Appendixes. In particular, we include an algebraic procedure for the explicit construction of the vector and scalar zero modes considered in the paper.

\section{Linear perturbations of a \rn  black hole} \label{lp}

This section contains all the material required for the proof of instability in Section \ref{sinter}.
We first review some basic facts on the maximal analytic extension of the \rn
solution to the Einstein-Maxwell equations (Section \ref{mae}),
and on the reduction of the linearized field equations around this
solution to decoupled 1+1 wave equations (Section \ref{rwz}).
Then in Section \ref{invs} we calculate the perturbed Riemann tensor invariants, to determine  the appropriate
boundary conditions at the singularity for the self-consistency of the perturbation method. In Section
\ref{asm} we review from \cite{prs, chandra} the factorization of the Regge-Wheeler and Zerilli hamiltonians,
and its connection to Chandrasekhar's ``algebraic special'' modes, which are central in the proof of instability that
follows.

\subsection{The \rn spacetime and its maximal analytic extension} \label{mae}

  The Reissner--Nordstr\"om  space-time metric
  \begin{equation} \label{rn}
  ds^2 = -f dt^2 + \frac{dr^2}{f} + r^2 (d \theta^2 + \sin ^2 \theta \; d \phi^2 ) =: g_{ab}dy^a dy^b + r^2 \hat g_{ij}
  dx^i dx^j ,
  \end{equation}
   is a warped product ${\cal N} \times_{r^2} S^2$ of a two dimensional Lorentzian ``orbit" manifold times
  a unit two sphere. The Maxwell field on this space-time is
\begin{equation} \label{max}
F = \frac{Q}{r^2} \; dt \wedge dr .
\end{equation} In (\ref{rn}),
  $f$ is the norm of the Killing vector $k^a = \p / \p t$,
\begin{equation} \label{f}
f = 1 - \frac{2M}{r} + \frac{Q^2}{r^2}  = \frac{(r -\ro)(r-\ri)}{r^2} ,
\end{equation}
the latter form being useful when $|Q| < M$, in which case the roots of $f$ are positive real numbers,
$0 < \ri < \ro$, and correspond to the horizon radii. It is useful to keep in mind the relation between
the alternative two-parameter descriptions of (\ref{rn})
\begin{eqnarray}
\ri &=& M - \sqrt{M^2-Q^2} ,  \;\; \ro = M + \sqrt{M^2-Q^2} \\
M &=& \frac{1}{2} (\ri+\ro) ,  \;\; Q^2 = \ri \ro.
\end{eqnarray}
 $k^a$ is timelike in the exterior ($r > \ro$) and
interior $0<r<\ri$ regions. As long as  we restrict to a region where $r \neq \ri, \ro$,  the coordinates
in (\ref{rn}) are appropriate. These coordinates become singular at $\ri$ and $\ro$, yet the \rn spacetime can be
extended through  the horizons, and new regions isometric
to I: $r > \ro$, II: $\ri < r < \ro$ and III: $0 <r < \ri$ arise ad infinitum, giving rise
to the Penrose diagram displayed in Figure 1. The stability of those regions isometric to III is the
subject of this paper.\\

\begin{figure}[h]
\centerline{\includegraphics[height=9cm]{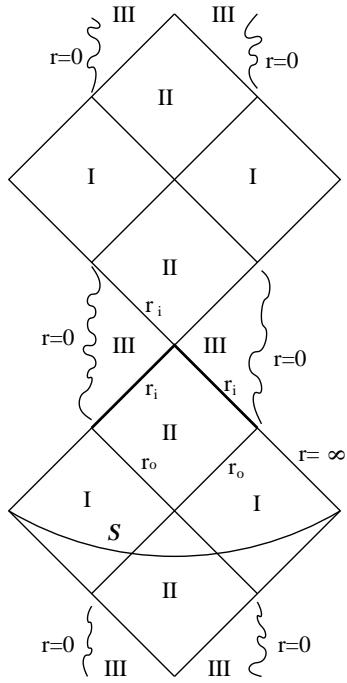}}
\caption{Penrose diagram for the maximal analytic extension of the \rn space-time. All regions labeled
 I  are isometric, and so are those labeled II, and III. Regions  I  and III are stationary,
and region  I  is known to be linearly stable. The function $r$ is globally defined, with $r_o<r$ in  I ,
$r_i<r<r_o$ in II , and $0<r<r_i$ in III. Also $r \to \infty$ at the conformal boundaries ${\cal J}^{\pm}$.
The $r=\ri$ horizon  drawn thicker is a Cauchy horizon for the initial data surface $S$, the ${\mathbb C}^{\infty}$
extension
 beyond it (in particular, the two copies of region III just above it being non unique, unless analyticity
of the metric is required.}
\end{figure}

Given a complete spacelike surface such as $S$ in Figure 1, a Cauchy horizon (thicker $\ri$ horizon
 in the figure) develops at $\ri$.
This is the boundary of the maximal domain of development of the data, and it is  connected to $S$ by
timelike curves of {\em finite} proper time. The solution  of the Einstein-Maxwell equations
is unique only up to the Cauchy horizon, and, although the spacetime is $C^{\infty}$ extensible beyond it -as shown
in Figure 1-  the extension is not determined by the data in $S$, and is non unique. This lack of predictability in
a classical theory of fields moved Penrose \cite{p1} to postulate what is known as the {\em strong cosmic censorship
conjecture}, according to which, for generic initial data in an appropriate class, the
maximal domain of development is inextensible (thus guaranteeing the preservation of predictability).
The idea that under more realistic assumptions than perfect spherical symmetry a Cauchy horizon would not
develop, is supported  by the finding that certain natural derivatives of a perturbation field diverge as
the Cauchy horizon is approached from region II  \cite{p2}, and by Israel and Poisson ``mass inflation'' model
\cite{ip}. An alternative, non perturbative approach was carried out by Dafermos \cite{daf}. In
\cite{daf},   spherical symmetric solutions
to the Einstein-Maxwell-scalar field system are studied. The (uncharged) scalar field was added
 to get around the uniqueness Birchoff theorems in the spherically symmetric case.
A characteristic problem is solved combining
  Reissner-Nordstr\"om data at the event horizon
with generic matching data
at the other null edge coming out the bifurcation sphere at $\ro$. It is shown that, generically,
the Hawking mass diverges at the Cauchy horizon, and thus
the spacetime fails to be $C^1$ extendible beyond it.\\

The results in this paper contribute to the idea that the extended spacetime depicted in Figure 1 is an irrelevant
solution of the Einstein-Maxwell equations. We show that a linear perturbation of the metric and Maxwell fields
in region III,
compactly supported away from the Cauchy horizon and the singularity, will grow exponentially in time, showing that
region III is in fact an unstable static solution of the Einstein-Maxwell equations.

\subsection{Linearized gravity around the \rn solution} \label{rwz}
The linearized Einstein-Maxwell equations  around (\ref{max})  have been analyzed by many authors, starting
with the papers by Regge, Wheeler and Zerilli
\cite{reggewheeler}, generalized to higher dimensional charged black holes with constant curvature horizons
 by Kodama and Ishibashi
  \cite{kodamaishibashi,ki}.
For polar (scalar in \cite{kodamaishibashi,ki}, here denoted $(+)$ following \cite{prs,chandra})
and axial (vector in \cite{kodamaishibashi,ki}, here denoted $(-)$ following \cite{prs,chandra})  modes
with harmonic number $\ell$ ($\ell= 2,3,...$), the
metric and electromagnetic perturbations  of a  Reissner--Nordstr\"om space-time
 can  be encoded in two
functions, $\Phi_{{\a}}^{\pm}(t,r), {\a}=1,2,$ that satisfy wave  equations \cite{prs,ki}
\begin{equation}
\label{eqp}
  0 =  \frac{\partial \Phi_ {\a}^{\pm}}{\partial t^2} - \frac{\partial \Phi_ {\a}^{\pm}}
{\partial x^2} + V^{\pm}_{\a} \Phi_ {\a}^{\pm} =: \frac{\partial \Phi_ {\a}^{\pm}}{\partial t^2}
+ {\cal H}^{\pm}_{\a} \Phi_ {\a}^{\pm}
\end{equation}
with potentials
 \begin{equation} \label{pots}
V^{\pm}_{\a} =
\pm  \beta_{{\a}} \frac{d f_{\a}}{d x} + \beta_{\a}{}^2 f_{\a}{}^2 + \kappa f_{\a}
 \end{equation}
where $\kappa = (\ell-1) \ell (\ell+1) (\ell+2)$,
$\beta_{\a} = 3M + (-1)^{\a} \sqrt{9 M^2+4 Q^2 (\ell-1)(\ell+2)}$,
\begin{equation}\label{fi}
f_{\a} = \frac{f}{r \beta_{\a} + (\ell -1)(\ell+2) r^2}
\end{equation}
with $f$ given in (\ref{f})
(equation (\ref{fi}) corrects a typo in \cite{doglepu}) and $x$ is a
``tortoise" coordinate, defined by $dx/dr = 1/f$. Eq. (\ref{eqp}) admits separation of variables
$\Phi_{\a}^{\pm}(t,r) = \exp(-i\w t) \psi_{\a}^{\pm}(r)$, leading to the Schr\"odinger like equation
\begin{equation}
\label{eqp2}
  \w^2  \psi_{\a}^{\pm} =   {\cal H}^{\pm}_{\a}  \psi_{\a}^{\pm}
\end{equation}
Unstable modes correspond to purely imaginary $\w$, and thus to negative eigenvalues
of the ``hamiltonian'' ${\cal H}$.\\

We are interested in the case $M > |Q|$ and $0 < r < r_i$, then
\begin{equation} \label{rs}
x =
r + \frac{\ro^2}{\ro-\ri} \ln \left( \frac{\ro-r}{\ro}\right) +  \frac{\ri^2}{\ri-\ro}
 \ln \left( \frac{\ri -r}{\ri} \right)
\end{equation}
where the integration constant was chosen  so that  $x$ ranges from zero to infinity as $r$ goes from zero to $\ri$. In these
limits
\begin{equation} \label{rsl}
x  \simeq \begin{cases} \frac{1}{3 \ri \ro}\; r^3 + \frac{\ri+\ro}{(2 \ri \ro)^2} \; r^4 + {\cal O} (r^5)&, r \to 0^+\\
\frac{\ri^2}{\ri-\ro}
 \ln \left( \frac{\ri -r}{\ri} \right) + ...   &, r \to {\ri}^- \end{cases}
\end{equation}

Note form (\ref{fi}) that $f_1$ has a singularity at
\begin{equation} \label{rc}
r_c =  \frac{\sqrt{9 M^2+4 Q^2 (\ell-1)(\ell+2)} -3M}{(\ell -1)(\ell+2)} ,
\end{equation}
then, from (\ref{pots}), we expect a singularity at $r_c$ in $V^{\pm}_1$.
However, the divergences from the different terms cancel out and  the vector potential $V^-_1$ is smooth at $r_c$.
This is not the case for the  scalar mode $V^+_1$, which
has a quadratic pole at $r_c$, with a positive coefficient.
The singularities at $r=0$ and $r=r_c$ have very different origins. The first one is due to the spacetime
singularity at this point, whereas the second one
 can be traced back to the definition of $\Phi^+_1$, which happens to be a singular function
of the metric and Maxwell field first order variations at this point (see, e.g., (\ref{s1})-(\ref{s3})),
this being the reason why
we refer to it as a ``kinematic" singularity. We should stress here that the way $\Phi^+_1$ is defined in terms
of the perturbed metric and Maxwell fields is crucial to disentangle the linearized Einstein-Maxwell
equations, and that $r_c$ happens to fall outside the domain of interest $r>\ro$ when
the stability of the {\em exterior} region is studied.
 Note that $r_c$ is a decreasing function of $\ell$ and an increasing function of $Q$. Thus,
for large
enough $\ell$ we have $0 < r_c <\ri$, and $V_1^+$ is singular in the inner region, the one that we study in this paper.
The ``safest'' mode is $\ell=2$,
for which $r_c > \ri$, and therefore the potential regular for $0 < r < \ri$, as long as $Q/M < \sqrt{7}/4 \simeq 0.66$.
For larger $Q/M$ rate,
 $r_c < r_i$  for every harmonic mode.\\
Fig. 2 depicts the $\ell=2$ scalar potentials $V_{\alpha}^+$ for some particular values of the parameters.
\begin{figure}[h]
\centerline{
\includegraphics[width=8cm,height=5cm,angle=0]{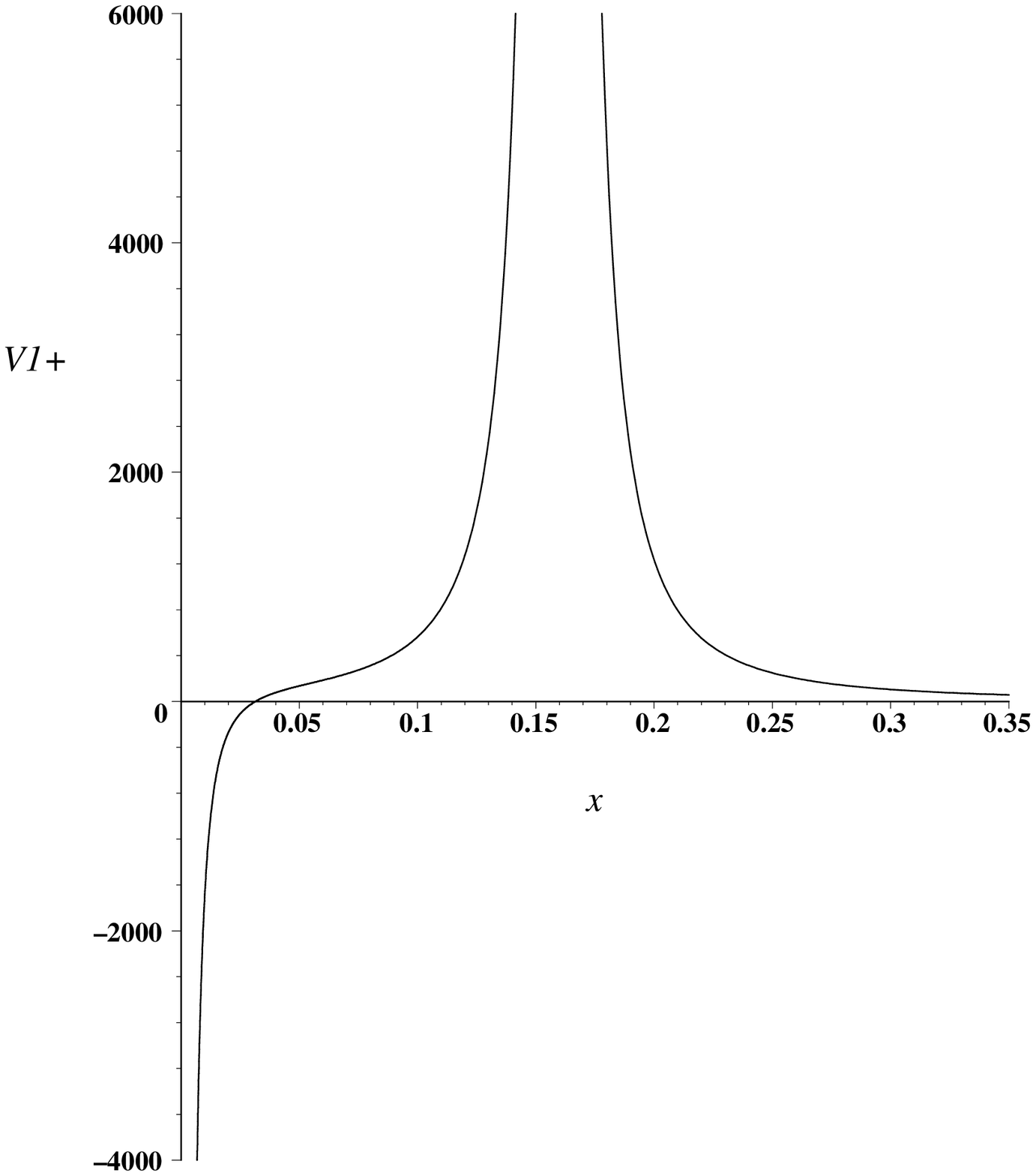} 
\hspace{1cm} \includegraphics[width=8cm,height=5cm,angle=0]{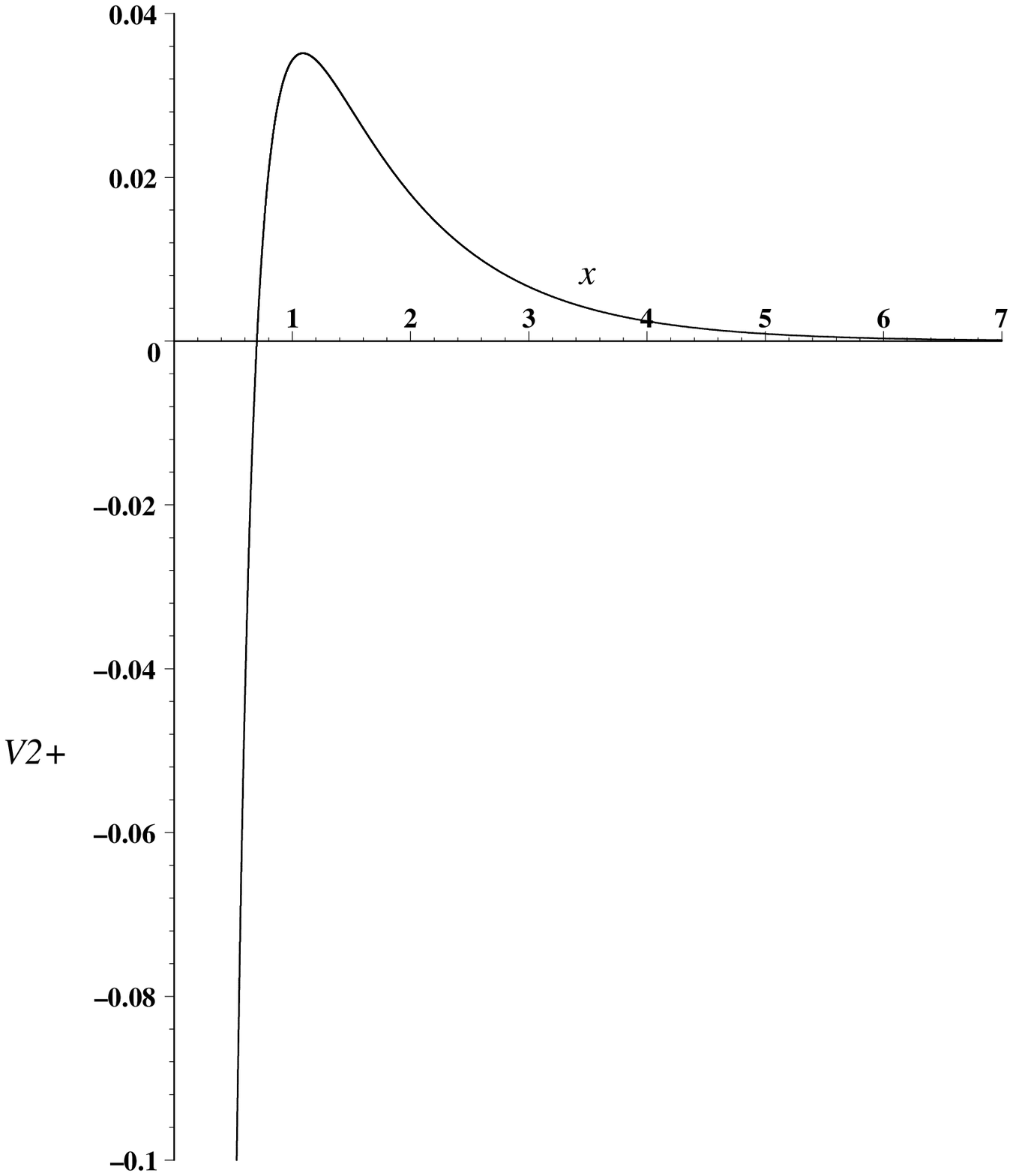}}
\caption{Scalar potentials $V_1^+$ (left) and $V_2^+$ (right) for $\ell=2, \ri=1, \ro=2$ plotted against $x$.
The $x$ range was chosen in each case to exhibit the relevant details, beyond these ranges the behavior is
that captured in equations (\ref{asymp})}
\end{figure}
The behaviour of the potentials near the spacetime singularity and the inner horizon is
\begin{equation} \label{asymp}
V_{\alpha}^+ \simeq  \begin{cases}  - \frac{2}{9 x{}^2} + ... & x \simeq 0 \\
C_{\alpha}^+ \exp \left(
 -\frac{(\ro-\ri)x}{\ri^2} \right) & x \to \infty \end{cases}
\end{equation}
where
\begin{equation}
C_{\alpha}^+ = \left( \frac{\ro-\ri}{\ri{}^4}\right) \left[ \frac{\kappa \ri{}^2 - \b_{\a}(\ro-\ri)}{\b_{\a} +  (\ell -1)(\ell+2) \ri} \right]
\end{equation}
The local solutions of the equation
\begin{equation}\label{evs}
{\cal H}_{\a}^+ \; \psi_{\a}^+ = -k^2 \; \psi_{\a}^+, 
\end{equation}
near the inner horizon and the singularity are (for both $\alpha=1$ and $\alpha=2$) of the form
\begin{equation}\label{asyms}
\psi_{\a}^+ \simeq  \begin{cases} A  \cos(\theta)  [ x{}^{1/3} \;  \sum_{n \geq 0} a_n^{(1)} x^{n/3} ] +
 A \; \sin(\theta) [ x{}^{2/3}  \sum_{n \geq 0} a_n^{(2)} x^{n/3} ]&
\text{ for } \;x
\simeq 0 \\ b_1 [\;  \exp (-k x) + ... ] +  b_2 [\;  \exp (k x) + ... ]
 & \text{ for } \; x \to \infty \end{cases}
\end{equation}
where we have set $a_0^{(1)}=1=a_0^{(2)}$.
The differential equation (\ref{evs}), rewritten using  $r$ as the independent variable, 
has a regular singular point at $r=0$, whose indicial equation has roots $1$ and $2$. The terms
between square brackets above are just the Frobenius series solution for this equation, written 
in terms of $x$ by inverting (\ref{rsl}) (thus the powers of $x^{1/3}$).
The two arbitrary
constants in front of them where parameterized with $A>0$ and $\theta \in [0,2\pi)$ for later convenience.
Similar expansions are made for local solutions of differential equations near the singularity 
at different points below.\\

Fig. 3 depicts the $\ell=2$ vector potentials $V_{\alpha}^-$ for the same parameter values as those in Fig. 2.\\
\begin{figure}[h]
\centerline{
\includegraphics[width=7cm,height=5cm,angle=0]{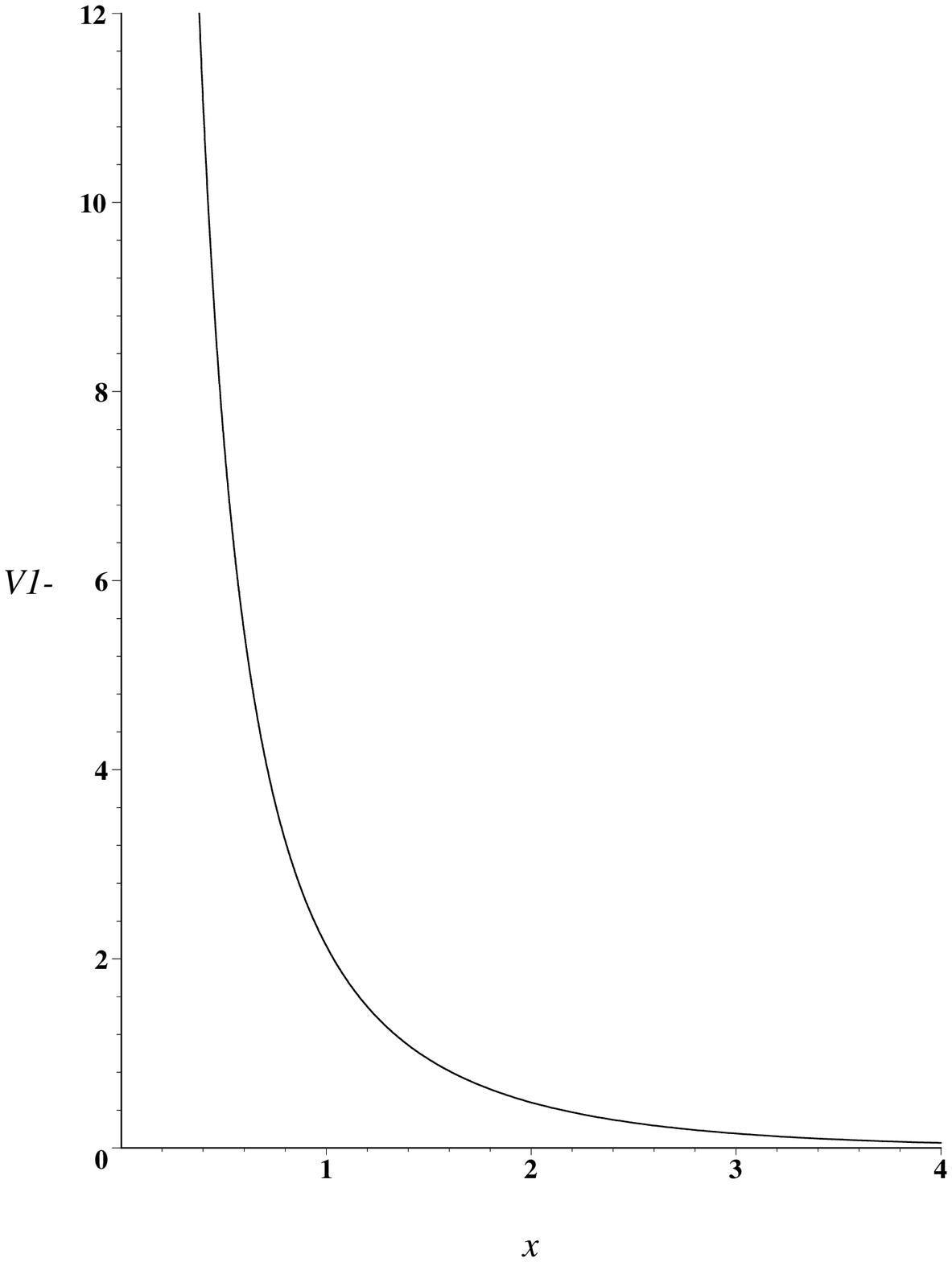} 
\hspace{1cm}
\includegraphics[width=7cm,height=5cm,angle=0]{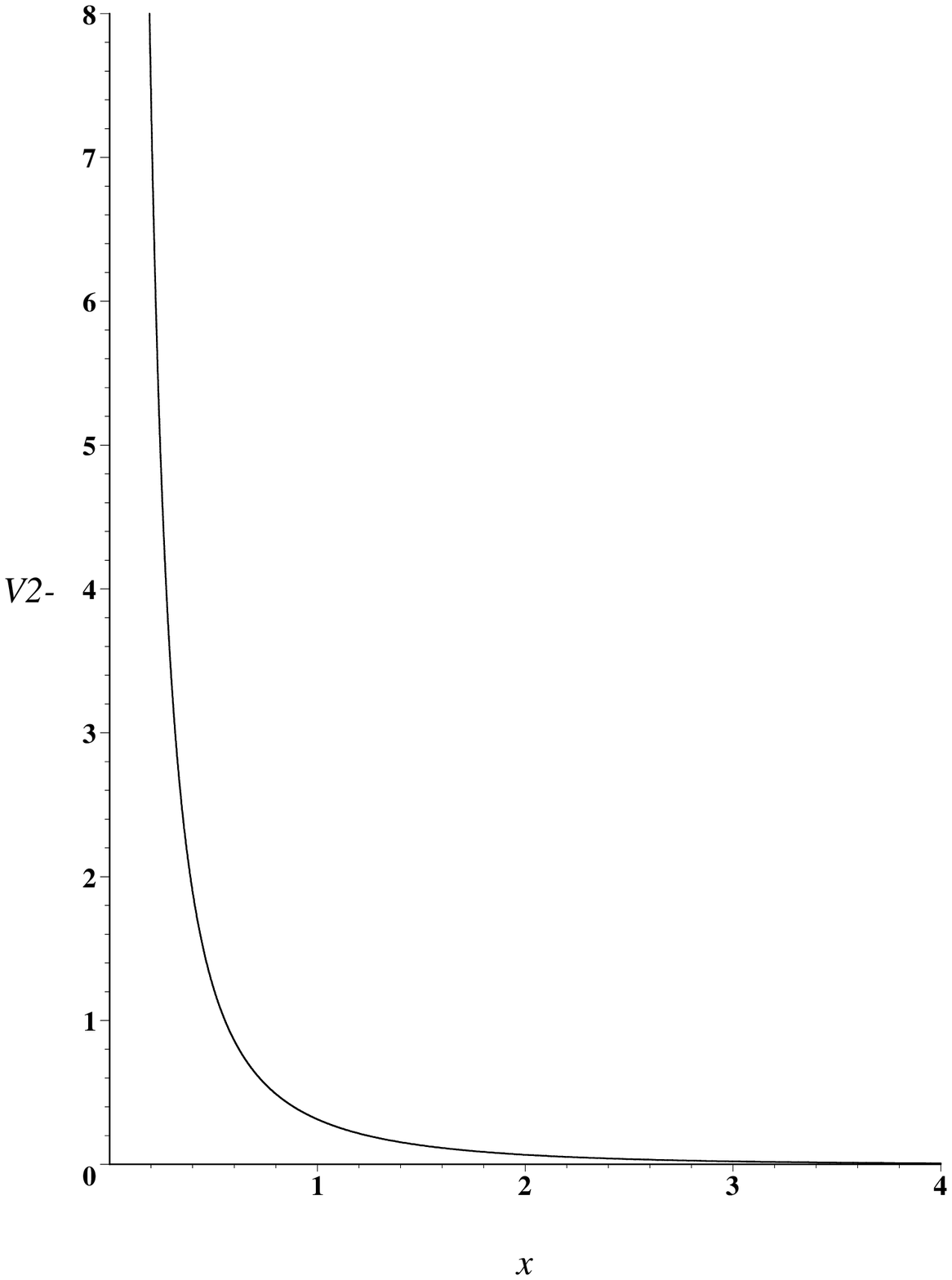}}
\caption{Vector potentials $V_1^-$ (left) and $V_2^-$ (right) for $\ell=2, \ri=1, \ro=2$ plotted against $x$.
}
\end{figure}

The behaviour of the potentials near the spacetime singularity and the inner horizon is
\begin{equation} \label{asymvp}
V_{\alpha}^- \simeq  \begin{cases}   \frac{4}{9 x{}^2} + ... & x \simeq 0 \\
C_{\alpha}^- \exp \left(
 -\frac{(\ro-\ri)x}{\ri^2} \right) & x \to \infty \end{cases}
\end{equation}
where
\begin{equation}
C_{\alpha}^- = \left( \frac{\ro-\ri}{\ri{}^4}\right) \left[ \frac{\kappa \ri{}^2 + \b_{\a}(\ro-\ri)}{\b_{\a} +  (\ell -1)(\ell+2) \ri} \right]
\end{equation}
The local solutions of the equation
\begin{equation}\label{evv}
{\cal H}_{\a}^- \; \psi_{\a}^- = - k^2 \; \psi_{\a}^- ,
\end{equation}
which correspond to a mode $\w=\pm ik$, are, for both $\alpha=1,2$, of the form
\begin{equation}\label{asymv}
\psi^-_{\a} \simeq  \begin{cases} A  \cos(\theta)  [ x{}^{-1/3} \;  \sum_{n \geq 0} a_n^{(1)} x^{n/3} ] +
 A \; \sin(\theta) [ x{}^{4/3}  \sum_{n \geq 0} a_n^{(2)} x^{n/3} ]&
\text{ for } \;x
\simeq 0 \\ b_1 [ \exp (-k x) + ... ] +  b_2 [ \exp (k x) + ... ]
 & \text{ for } \; x \to \infty \end{cases}
\end{equation}
where we have set $a_0^{(1)}=1=a_0^{(2)}$.

\subsection{Consistency of the linearized analysis}  \label{invs}

In the linearized approach, a solution $g_{ab}, A_a$ of the Einstein-Maxwell equations is replaced with
a ``perturbed'' metric and electromagnetic field potential $g_{ab}+ \e h_{ab}, A_a + \e B_a$, and the
field equations are then required to hold to first order in $\e$.
Given that the background solution we are interested in, region III of the \rn spacetime, has a curvature
singularity as $r \to 0^+$, the  perturbation treatment will certainly be inconsistent if we find that
the first order correction to a  perturbed divergent curvature scalar diverges faster than the unperturbed piece
in the $r \to 0^+$ limit,
since  in this  case the notion of a ``uniformly small metric perturbation'' is lost.
  This is why  the first order piece of the Kretschmann or some
other invariant is usually computed.
 Here we take a systematic approach
to make sure  that {\em none } of the  algebraic invariant made out
of the Riemman tensor acquires a correction diverging faster than the corresponding background metric invariant.\\
Any algebraic polynomial invariant of the Riemann tensor $R_{abcd}$ can be written as a  polynomial on a set of basic
invariants, the basic invariants being generically subject to syzygies (constraints).
The basic invariants are more   compactly written
in terms of the Ricci tensor $R_{ab} := R^c{}_{a c b}$, the Ricci scalar $R=R^a{}_a$, the trace free Ricci
tensor $S_{ab}=R_{ab} -
g_{ab} R/4$,  the Weyl tensor
$$C_{abcd} := R_{abcd} - \frac{2}{n-2}(g_{a[c}S_{d]b} - g_{b[c}S_{d]a})  - \frac{2}{n(n-1)} R g_{a[c} g_{d]b} $$
and its dual $C^*_{abcd} :=\frac{1}{2}\epsilon_{abef} C^{ef}{}_{cd}$, or just using the Ricci and Weyl spinors
$\Phi_{AB\dot{A}\dot{B}}, \Psi_{ABCD}$. In the case of spacetimes with a perfect fluid or a Maxwell field,
the basic invariants are those given in Table I (from \cite{carmi}.) \vspace{1cm}\\

\renewcommand{\baselinestretch}{1.5}\normalsize
\begin{tabular}{rl}\hline
&Table I:  Basic Riemann tensor invariants for perfect fluid or Maxwell field spacetimes   \\  \hline
$R$ & $:= R^a{}_a$ \\
$R_1$ & $:= \Phi_{AB\dot{A}\dot{B}}
  \Phi^{AB\dot{A}\dot{B}}= \frac{1}{4} S^a{}_b S^b{}_a$  \\
$R_2$ & $:= \Phi^A{}_B{}^{\dot{A}}{}_{\dot{B}}
  \Phi^B{}_C{}^{\dot{B}}{}_{\dot{C}}
  \Phi^C{}_A{}^{\dot{C}}{}_{\dot{A}}= -\frac{1}{8} S^a{}_b S^b{}_c S^c{}_a$   \\
$R_3$& $:= \Phi^A{}_B{}^{\dot{A}}{}_{\dot{B}}
  \Phi^B{}_C{}^{\dot{B}}{}_{\dot{C}}
  \Phi^C{}_D{}^{\dot{C}}{}_{\dot{D}}
  \Phi^D{}_A{}^{\dot{D}}{}_{\dot{A}}= \frac{1}{16} S^a{}_b S^b{}_c S^c{}_d
  S^d{}_a$       \\
 $w_1$ & $:= \Psi_{ABCD}\Psi^{ABCD} = \frac{1}{8} ( C_{abcd}
  + i C^*_{abcd} ) C^{abcd}$       \\
$w_2$ & $:= \Psi^{AB}{}_{CD}\Psi^{CD}{}_{EF}
  \Psi^{EF}{}_{AB}= -\frac{1}{16} ( C_{ab}{}^{cd}
  + i C^*_{ab}{}^{cd} ) C_{cd}{}^{ef}
  C_{ef}{}^{ab}$      \\
 $m_1$ & $:= \Psi_{ABCD} \Phi^{AB{\dot{A}}{\dot{B}}}
  \Phi^{CD}{}_{{\dot{A}}{\dot{B}}}= \frac{1}{8} S^{ab} S^{cd} ( C_{acdb}
  + i C^*_{acdb})$   \\
 $m_2$& $:= \Psi_{ABCD} \Psi^{AB}{}_{EF}
  \Phi^{CD{\dot{A}}{\dot{B}}}
  \Phi^{EF}{}_{{\dot{A}}{\dot{B}}}$   \\
$m_3$   & $:= \Psi_{ABCD}
  \bar{\Psi}_{\dot{A}\dot{B}\dot{C}\dot{D}}
  \Phi^{AB\dot{A}\dot{B}}
  \Phi^{CD\dot{C}\dot{D}}$    \\
$m_4$ & $:=\Psi_{ABCD}
  \bar{\Psi}_{\dot{A}\dot{B}\dot{C}\dot{D}}
\Phi^{AB\dot{C}\dot{E}} \Phi^{CE\dot{A}\dot{B}} \Phi^D{}_E{}^{\dot{D}}{}_{\dot{E}}$    \\
 $m_5$ & $:= \Psi_{ABCD}\Psi^{CDEF}
  \bar{\Psi}^{\dot{A}\dot{B}\dot{E}\dot{F}}
  \Phi^{AB}{}_{\dot{A}\dot{B}}
  \Phi_{EF\dot{E}\dot{F}}$    \\
\hline
\end{tabular}
\renewcommand{\baselinestretch}{1}\normalsize
{} \vspace*{1cm}\\

In the Maxwell case, $R=R_2 =0$, and the syzygies  among the remaining invariants are \cite{carmi}
\begin{equation} \label{const}
R_1{}^2 = 4 R_3, \;\; m_4=0, \;\; m_1 \bar m _2 = R_1 \bar m _5, \;\;
m_2 \bar m _2 m_3 = R_1 m_5 \bar m_5
\end{equation}
For \rn $R_1$ and $m_2$ do not vanish, then the  syzygies imply that, as long as
  $R_1,w_1,w_2,m_1$ and $m_2$ behave properly (correction does not diverge faster than unperturbed term),
the same will happen to any curvature invariant, of any degree.
Using references \cite{reggewheeler,kodamaishibashi}, we  have reconstructed the perturbed metric for each
mode, and then calculated the perturbed invariants with the help of the grtensor symbolic manipulation package \cite{grtensor}.
For the vector (axial) modes we could reduce all expressions, by repeatedly applying  (\ref{eqp}), to
\begin{eqnarray} \nonumber
R_1 &=& \frac{Q^4}{r^8} \\  \nonumber
w_1 &=& \frac{6(Q^2-Mr)^2}{r^8} + i \epsilon \; \frac{6(Q^2-Mr) {\cal Z}}{r^{8}} \;  Y_{\ell m}(\theta,\phi)\\ \label{iv1}
w_2 &=&  \frac{6 (Q^2-Mr)^3}{r^{12}} + i \epsilon \; \frac{9(Q^2-Mr)^2  {\cal Z}}{r^{12}} \; Y_{\ell m}(\theta,\phi)\\  \nonumber
m_1 &=&  \frac{2 (Q^2-Mr) Q^4}{r^{12}} +  i \epsilon \; \frac{Q^4  {\cal Z}}{r^{12}} \;  Y_{\ell m}(\theta,\phi)\\  \nonumber
m_2 &=&   \frac{4 (Q^2-Mr)^2 Q^4}{r^{16}} - i \epsilon \; \frac{4 (Q^2-Mr) Q^4  {\cal Z}}{r^{16}} \; Y_{\ell m}(\theta,\phi)
\end{eqnarray}
where
\begin{equation} \label{iv2}
{\cal Z} := \left[  \frac{ \kappa  (\beta_2 r -4Q^2) \psi_2^-}{ 2(\b_2 - \b_1)}
 + \frac{2 Q (\beta_2 + (\ell+2)(\ell-1) r) \ell (\ell+1) \psi_1^-}{\b_2-\b_1}   \right]
\end{equation}
From (\ref{asymv}), (\ref{iv1}) and (\ref{iv2}),
  it is clear that an unstable vector mode gives an  inconsistent perturbation unless $a_0^{(1)}=b_2=0$ in (\ref{asymv}).
It is only in this case that the perturbation can be uniformly bound in the whole of region III.\\

For the scalar modes, the invariants can be expressed entirely in terms of the Zerilli field {\em and its first
$r-$derivative} but we were not able to reduce the resulting expressions to a reasonably compact form,
except for $R_1$, for which we found that
\begin{equation} \label{dr1}
R_1 = \frac{Q^4}{r^8} \left[ 1 - 4 \epsilon \left( f \frac{\p \psi_1^+}{\p r} - f\frac{\p \chi_1^+}{\p r} \frac{\psi_1^+}{\chi_1}
\right)Y_{\ell m}(\theta,\phi) - 4 \epsilon \left( f \frac{\p \psi_2^+}{\p r} - f \frac{\p \chi_2^+}{\p r} \frac{\psi_2^+}{\chi_2}
\right) Y_{\ell m}(\theta,\phi)\right],
\end{equation}
where the $\chi_{\a}^+$ is one of Chandrasekhar's algebraic special modes, introduced in the following Section (see equation
(\ref{chi})). The above formula will turn out to be very useful in the following sections.\\

\subsection{Factorization of Zerilli's Hamiltonian and algebraic special modes}\label{asm}

Chandrasekhar's algebraic special modes (ASM) are solutions of equations (\ref{evs}), (\ref{evv}) with real $k$,
 i.e., unstable
modes of the perturbation equations. These modes do not satisfy appropriate boundary conditions as linear
perturbations of region  I  of the \rn black hole, in agreement with the fact that this region is linearly stable.
However, some ASM were shown in \cite{doglepu,cardoso} to satisfy appropriate boundary conditions as
perturbations of the \rn naked singularity, showing that this spacetime is unstable.
In this Section we consider algebraic special modes as perturbations of region III
of a \rn black hole, and analyze their behaviour near the singularity and the inner horizon.
Following \cite{prs}, we introduce $\chi^{\pm}_{\a}$ defined by
\begin{equation} \label{chis}
\frac{d}{dx} \ln \chi^{\pm}_{\a} = \pm \left( \beta_{\a} f_{\a} + \frac{\kappa}{2 \beta_{\a}} \right) ,
\end{equation}
the general solution of this equation being an irrelevant  constant times
\begin{equation} \label{chi}
\chi_{\a}^{+} =  \frac{M r \exp \left( \frac{\kappa x}{2 \beta_{\a}} \right) }{ \beta_{\a} +(\ell-1)(\ell+2)r}
\hspace{2cm} \chi_{\a}^{-} = M \exp \left( \frac{-\kappa x}{2 \beta_{\a}} \right)
 \frac{ ( \beta_{\a} +(\ell-1)(\ell+2)r)}{r}.
\end{equation}
In terms of the functions $\chi_{\a}^{\pm}$, equation (\ref{eqp2}) reads
\begin{equation} \label{eas}
\frac{1}{\psi_{\a}^{\pm}} \: \frac{d^2 \psi_{\a}^{\pm}}{dx{}^2} + \left[  \w^2 + \frac{\kappa^2}{4 (\beta_{\a})^2} \right]  =
\frac{1}{\chi_{\a}^{\pm}} \: \frac{d^2 \chi_{\a}^{\pm}}{dx{}^2} ,
\end{equation}
which can easily be integrated if
\begin{equation} \label{as}
\w = \pm i k_{\a} , \;\; k_{\a} :=  \frac{\kappa}{2 |\beta_{\a}|} > 0
\end{equation}
Equation (\ref{as}) defines the {\em algebraic special modes}, (ASM) which give
 unstable solutions  $\Phi_{\a}^{\pm}(t,r)=\exp(k_{\a} t) \zeta_{\a}^{\pm}(r)$
of the perturbation equations, with $\zeta_{\a}^{\pm}$ a
solution of (\ref{eas}) when the term between brackets vanishes:
\begin{equation} \label{modes}
\zeta_{\a}^{\pm} = A_{\a}^{\pm} \; \chi_{\a}^{\pm} + B_{\a}^{\pm} \;  \tau_{\a,R^*}^{\pm} , \hspace{.4cm}
\tau_{\a,R^*}^{\pm} := \chi_{\a}^{\pm}
\int^{x}_{R^*} \frac{M dx}{[ \chi_{\a}^{\pm}(x) ]^2} .
\end{equation}
Note that a change of choice of $R^*$ amounts to adding a constant times $\chi$ to $\tau$,
and thus the constants $A$ and $B$ in (\ref{modes}) are unambiguously defined only after
$R^*$ has been chosen.\\
Alternatively, ASM can be obtained from the factorization property of the Hamiltonians ${\cal H}_{\a}^{\pm}$
\begin{eqnarray} \label{fsh}
{\cal H}_{\a}^{+} &=& {\cal A}_{\a} {\cal B}_{\a} - \left( \frac{\kappa}{2 \beta_{\a}} \right)^2\\
{\cal H}_{\a}^{-} &=& {\cal B}_{\a} {\cal A}_{\a} - \left( \frac{\kappa}{2 \beta_{\a}} \right)^2 , \label{fvh}
\end{eqnarray}
where
\begin{eqnarray}
{\cal A}_{\a} &:=& \frac{\partial}{\partial x} + \left( \beta_{\a} f_{\a} + \frac{\kappa}{2 \beta_{\a}} \right)\\
{\cal B}_{\a} &:=& -\frac{\partial}{\partial x} + \left( \beta_{\a} f_{\a} + \frac{\kappa}{2 \beta_{\a}} \right).
\end{eqnarray}
$\zeta_{\a}^{+}$ span de kernel of ${\cal A}_{\a} {\cal B}_{\a}$, with $\chi_{\a}^{+}$ in ker ${\cal B}_{\a}$, whereas
 $\zeta_{\a}^{-}$ span de kernel of ${\cal B}_{\a} {\cal A}_{\a}$, with $\chi_{\a}^{-}$ in ker ${\cal A}_{\a}$.\\

\subsubsection{Algebraic special vector modes}

The asymptotic behaviour of  $\chi_{\a}^{-}$
near the spacetime singularity and the inner horizon is
\begin{equation} \label{avv}
 \chi_{\alpha}^{-} \simeq  \begin{cases}   \frac{M \beta_{\a}}{(3 \ri \ro )^{1/3}} \; x{}^{-1/3}   + ... & x \simeq 0 \\
\frac{M}{\ri} \left[ \beta_{\a} +(\ell-1)(\ell+2)r_i \right] \exp \left( -\frac{\kappa x}{2 \beta_{\a}} \right) +
 ... & x \to \infty
\end{cases}
\end{equation}

 The vector $\tau$ modes are
\begin{equation} \label{tauvect}
\tau_{\a,R^*}^{-} :=  \left[ \frac{ \beta_{\a} +(\ell-1)(\ell+2)r}{ r \exp \left( \frac{\kappa x}{2 \beta_{\a}} \right) } \right]
\int^{x}_{R^*} \frac{ r^2 \exp \left( \frac{\kappa x}{\beta_{\a}} \right) dx}{[ \beta_{\a} +(\ell-1)(\ell+2)r]^2 }.
\end{equation}
Since $\beta_1 < 0$ and $0 < \beta_2$, for
$\a=2$ the integral in (\ref{tauvect})  near  $x = \infty$ ($r=\ri$) diverges, we can give  $R^*$ any
 {\em finite}
 value, and the asymptotic behaviour will depend on whether $R^* =0$ or not:
 \begin{equation}
 \tau_{\alpha=2, R^*=0}^{-} \simeq  \begin{cases}   x{}^{4/3}   + ... & x \simeq 0 \\
 \exp \left( \frac{\kappa x}{2 \beta_{2}} \right) + ... & x \to \infty
\end{cases}
\end{equation}
\begin{equation}
 \tau_{\alpha=2, R^* \neq 0}^{-} \simeq  \begin{cases}   x{}^{-1/3}   + ... & x \simeq 0 \\
 \exp \left( \frac{\kappa x}{2 \beta_{2}} \right) + ... & x \to \infty
\end{cases}
\end{equation}
Since those type-2 vector ASM that grow slower than $r^0$ as $r \to 0^+$ blow up at the inner horizon,
it follows from the analysis of invariants in the previous subsection (equations (\ref{iv1})-(\ref{iv2})), that
a pure  mode $\Phi_{2}^- = e^{k_2 t} \; \zeta_2^- =  e^{k_2 t} \; [
A_{2}^{-} \; \chi_{2}^{-} + B_{2}^{-} \;  \tau_{2,R^*}^{-}]$ cannot
 be consistently treated as a first order perturbation
on the entire \rn region III.\\
If $\a=1$ the integral in (\ref{tauvect})  near infinity converges, thus, we can give  $R^*$ any
 {\em finite}
 value, or take $R^* = \infty$. The asymptotic behaviour will be
 \begin{equation} \label{avt}
 \tau_{\alpha=1, R^*=0}^{-} \simeq  \begin{cases}   x{}^{4/3}   + ... & x \simeq 0 \\
 \exp \left( \frac{-\kappa x}{2 \beta_{1}} \right) + ... & x \to \infty
\end{cases}
\end{equation}
\begin{equation}
 \tau_{\alpha=1, R^* \neq 0}^{-} \simeq  \begin{cases}   x{}^{-1/3}   + ... & x \simeq 0 \\
 \exp \left( -\frac{\kappa x}{2 \beta_{1}} \right) + ... & x \to \infty
\end{cases}
\end{equation}
\begin{equation}
 \tau_{\alpha=1, R^*=\infty}^{-} \simeq  \begin{cases}   x{}^{-1/3}   + ... & x \simeq 0 \\
 \exp \left( \frac{\kappa x}{2 \beta_{1}} \right) + ... & x \to \infty
\end{cases}
\end{equation}
It follows again from equations (\ref{iv1})-(\ref{iv2})  that
a pure AS mode $\Phi_{1}^- = e^{k_1 t} \; \zeta_1^- =  e^{k_1 t} \; [
A_{1}^{-} \; \chi_{1}^{-} + B_{1}^{-} \;  \tau_{1,R^*}^{-}]$ cannot
 be consistently treated as a first order perturbation
in the entire \rn region III.\\

Note that  some care is
required when interpreting equation (\ref{tauvect}) for the $\alpha=1$ vector mode,
due to  the singularity of the integrand at $r_c$. Take, e.g., the case $R^* =\infty$.
The one form under the integral sign  in
\begin{equation} \label{v1}
\tau_{1,\infty}^{-} :=   \left[ \frac{ (r_c-r) }{(\ell-1)(\ell+2) r \; \exp \left( \frac{\kappa x}{2 \beta_{1}}\right)}  \right]
\int_{x}^{\infty} \frac{ r^2  }{(r-r_c)^2 } \exp \left( \frac{\kappa x}{\beta_{1}} \right) \; dx.
\end{equation}
can be written as
$\left[ \frac{A}{(r-r_c)^2}+ \frac{B}{r-r_c} + Z(r) \right] dr$, $Z(r)$ the regular function obtained by subtracting the second
and first order poles. The
 integration constants at both sides of $r_c$ can then be adjusted such that
$$\tau_{1,\infty}^{-} = \frac{A - B (r-r_c) \ln |r-r_c| + \left( B \ln |\ri-r_c|-\frac{A}{\ri-r_c}  \right) (r-r_c) +
(r-r_c)\int_{r}^{\ri} Z(r) dr }{(\ell-1)(\ell+2) r \; \exp \left( \frac{\kappa x}{2 \beta_{1}}\right)}.$$
This is well defined across $r_c$.

\subsubsection{Algebraic special scalar modes}

The asymptotic behaviour of  $\chi_{\a}^{+}$
near the spacetime singularity and the inner horizon is
\begin{equation} \label{avs}
 \chi_{\alpha}^{+}   \simeq  \begin{cases} \frac{M}{\beta_{\a}} (3 \ri \ro )^{1/3} \; x{}^{1/3}   + ... & x \simeq 0 \\
\frac{M r_i  }{ \beta_{\a} +(\ell-1)(\ell+2)r_i} \; \exp \left( \frac{\kappa x}{2 \beta_{\a}} \right) + ... & x \to \infty \end{cases}
\end{equation}
The asymptotic behaviour of $\tau_{\a,R^*}^{\pm}$  depends on the choice of  $R^*$ in
\begin{equation} \label{taus}
\tau_{\a,R^*}^{+} :=  \left[ \frac{ r \exp \left( \frac{\kappa x}{2 \beta_{\a}} \right) }{ \beta_{\a} +(\ell-1)(\ell+2)r} \right]
\int^{x}_{R^*} \frac{[ \beta_{\a} +(\ell-1)(\ell+2)r]^2 dx}{ r^2 \exp \left( \frac{\kappa x}{\beta_{\a}} \right)} .
\end{equation}
If $\a=1$ the integral (\ref{taus}) diverges near infinity, then   $R^*$ is restricted to
 {\em finite} values, and
\begin{equation} \label{as10}
 \tau_{\alpha=1, R^*=0}^{+} \simeq  \begin{cases}   x{}^{2/3}   + ... & x \simeq 0 \\
 \exp \left(- \frac{\kappa x}{2 \beta_{1}} \right) + ... & x \to \infty
\end{cases}
\end{equation}
\begin{equation} \label{as1n}
 \tau_{\alpha=1, R^* \neq 0}^{+} \simeq  \begin{cases}   x{}^{1/3}   + ... & x \simeq 0 \\
 \exp \left( -\frac{\kappa x}{2 \beta_{1}} \right) + ... & x \to \infty
\end{cases}
\end{equation}
If $\a=2$ there are three possibilities:
\begin{equation} \label{as20}
 \tau_{\alpha=2, R^*=0}^{+} \simeq  \begin{cases}   x{}^{2/3}   + ... & x \simeq 0 \\
 \exp \left( \frac{\kappa x}{2 \beta_{2}} \right) + ... & x \to \infty
\end{cases}
\end{equation}
\begin{equation} \label{as2n}
 \tau_{\alpha=2, R^* \neq 0}^{+} \simeq  \begin{cases}   x{}^{1/3}   + ... & x \simeq 0 \\
 \exp \left( \frac{\kappa x}{2 \beta_{2}} \right) + ... & x \to \infty
\end{cases}
\end{equation}
\begin{equation} \label{as2i}
 \tau_{\alpha=2, R^*=\infty}^{+} \simeq  \begin{cases}   x{}^{1/3}   + ... & x \simeq 0 \\
 \exp \left( -\frac{\kappa x}{2 \beta_{2}} \right) + ... & x \to \infty
\end{cases}
\end{equation}
The only scalar AS modes that behave appropriately near the inner horizon are $\chi_{\alpha=1}^+$ and
$\tau_{\alpha=2, R^*=\infty}^{+}$. Since the integrals defining the latter one are non elementary, we
proceed with $\chi_{\alpha=1}^+$, for which an explicit reconstruction of the perturbed metric and invariants
is relatively simple. No invariant that is trivial at order zeroth develops a first order correction.
The non zero invariants to first order for the $\chi_1^+$ perturbation are
\begin{eqnarray} \label{i1}
R_1 &=& \frac{Q^4}{r^8} \\
w_1 &=& \frac{6(Mr-Q^2)^2}{r^8} -  \epsilon \; \left[ \frac{3 M \beta_2 \; \ell (\ell+1) \; (Mr-Q^2) }{2 \beta_1 r^7} \right] \; Y_{\ell m}(\theta,\phi)\;  e^{-\frac{\kappa}{2 \beta_1} (t-x)}
\\
w_2 &=& - \frac{6 (Mr-Q^2)^3}{r^{12}} + \epsilon \; \left[ \frac{9 M \beta_2 \; \ell (\ell+1)  \; (Mr-Q^2)^2 }{4 \beta_1 r^{11}} \right] \; Y_{\ell m}(\theta,\phi)\;  e^{-\frac{\kappa}{2 \beta_1} (t-x)}\\
m_1 &=& - \frac{2 (Mr-Q^2) Q^4}{r^{12}} + \epsilon \; \left[ \frac{M Q^ 4 \beta_2 \; \ell (\ell+1) }{4 \beta_1 r^{11}} \right] \; Y_{\ell m}(\theta,\phi)\;  e^{-\frac{\kappa}{2 \beta_1} (t-x)}\\
m_2 &=&  m_3 = \frac{4 (Mr-Q^2)^2 Q^4}{r^{16}} - \epsilon \; \left[ \frac{M Q^ 4 \beta_2 \; \ell (\ell+1) \; (Mr-Q^2)}{ \beta_1 r^{15}} \right] \; Y_{\ell m}(\theta,\phi)\;  e^{-\frac{\kappa}{2 \beta_1} (t-x)}
\label{i6}
\end{eqnarray}
This algebraic special mode satisfies all our requirements for a self consistent first order
formalism in this singular spacetime. Thus, we will restrict to type one scalar perturbations
from now on. $\chi_1^+$ is an example of a spatially uniformly bounded perturbation that
grows exponentially in time, and thus a signal of a gravitational instability. In the following Section,
we will show that this mode can be excited by a generic perturbation that is initially compactly supported within
region III. The treatment will  follow closely the case of the Schwarzschild
naked singularity treated in \cite{nm}.

\section{Proof of the linear instability of the inner static region} \label{sinter}

The linear {\em instability} of a static spacetime is established once an
unstable mode is found. We have shown in the previous section that, out of the
four modes existing for every harmonic pair $(\ell,m)$, the type=1 scalar mode
admits an unstable solution to the linearized Einstein-Maxwell equations -Chandrasekhar's AS mode-
that can
consistently be treated to first order in the whole domain of region III, $0 < r < \ri$.
The purpose of this section is to show how generic initial data with compact support in region III
excites this mode. Although this problem is trivial for perturbations in region  I,
it exhibits a number of unexpected difficulties when dealing with  perturbations of region III.
Zerilli's wave equation for this mode:
\begin{equation}
\label{se1}
  0 =  \frac{\partial \Phi_ {1}^{+}}{\partial t^2} - \frac{\partial \Phi_ {1}^{+}}
{\partial x^2} + V^{+}_{1} \Phi_ {1}^{+} =: \frac{\partial \Phi_ {1}^{+}}{\partial t^2}
+ {\cal H}^+_1 \Phi_ {1}^{+}
\end{equation}
has a potential with a singularity at $r_c$ given in equation (\ref{rc}), that generically
falls in region III (see left panel in Figure 2). The origin of this singularity
(a second order pole) in $V^+_1$ can be traced back to the definition of
$\Phi_1^+$ \cite{reggewheeler,kodamaishibashi}.
The first order variation of the electromagnetic field has
\begin{eqnarray}
\d F_{r \theta} &=&  \frac{\p {\cal A}}{\partial t} \frac{\p Y_{\ell m}}{\p \theta} \; f^{-1} , \\
\d F_{t \theta} &=&  \frac{\p {\cal A}}{\partial r} \frac{\p Y_{\ell m}}{\p \theta} \; f ,
\end{eqnarray}
then  ${\cal A}$ must be smooth for $\delta F$  to be smooth.
Since
\begin{equation}\label{s3}
{\cal A} = \frac{\beta_2 (r-r_c)}{8 Q r} \; \Phi_1^+(t,r),
\end{equation}
we conclude that, for generic smooth perturbations, $\Phi_1^+$ has a first order pole at $r=r_c$, and
can be Laurent expanded around $r_c$ as
\begin{equation} \label{series}
\Phi_1^+ = \sum_{k \geq -1} c_k (r-r_c)^k.
\end{equation}
The variation of $g_{tr}$ can be simplified to the form \cite{reggewheeler,kodamaishibashi}
\begin{equation} \label{gtr}
\d g_{tr} =  Y_{\ell m} \;\frac{\p}{\p t} \left[ r \frac{\p \Phi_1^+}{\p r} + {\cal B}\;\; \Phi_1^+ \right]  ,
\end{equation}
where
\begin{multline}
{\cal B} ={\frac { \left(  \left( \ell-1 \right)
 \left( \ell+2 \right) {r}^{4}-3\,M \left( -3+{\ell}^{2}+\ell \right) {r}^{3}+
 \left( 2\, \left( \ell-1 \right)  \left( \ell+2 \right) {Q}^{2}-12\,{M}^{2}
 \right) {r}^{2}+13\,{Q}^{2}Mr-4\,{Q}^{4} \right) }{ \left( {r}^{2}-2\,Mr+{Q}^{2} \right)  \left(  \left( \ell-1
 \right)  \left( \ell+2 \right) {r}^{2}+6\,Mr-4\,{Q}^{2} \right) }}\\
+ {\frac { r  \sqrt{9 M^2+4 Q^2 (\ell-1)(\ell+2)}}{   \left(  \left( \ell-1 \right)  \left( \ell+2 \right) {r}^{2}
+6\,Mr-4\,{Q}^{2} \right) }}
\end{multline}
It can be checked that  the $(r-r_c)^{-2}$ coefficient of
the series expansion of  (\ref{gtr}) vanishes. The $(r-r_c)^{-1}$ coefficient
will vanish if
\begin{equation} \label{cond}
\frac{c_0}{c_{-1}} = - \frac{(\ell-1)^2(\ell+2)^2(2M-\beta_1) }{\beta_1\left(2(\ell-1)
(\ell+2)M +(\ell^2+\ell+2)\beta_1\right)}
\end{equation}
It turns out that (\ref{series}) and (\ref{cond}) are not only necessary but also sufficient conditions
for the perturbations of the metric and electromagnetic fields to be smooth. Thus, we have proved that 
the   Zerilli functions $\Phi_1^+$ corresponding to smooth type-1 scalar perturbations are those
admitting, at any fixed time, a Laurent  expansion (\ref{series}) around $r_c$ satisfying the condition (\ref{cond}).\\
We  also need to check that the perturbation does not change the character of the singularity.
As explained in the previous section, this guarantees the self-consistency of the linearized theory. 
It then follows from equation (\ref{dr1}) that we must demand that, for some positive $\d$  and $N$
\begin{equation} \label{cond0}
 \Big | f \frac{\p \Phi_1^+}{\p r} - f \frac{\p \chi_1}{\p r} \frac{\Phi_1^+}{\chi_1} \Big | \leq N,  \;\;
\text{ if } \;\; 0 < r < \d
\end{equation}
\noindent
We will show in the following Section that this condition is also sufficient to assure that
 all the  invariants behave properly.
Thus, we arrive at:\\

\noindent
{\bf Lemma 1}
In order that the metric and electromagnetic scalar type-1 perturbations be smooth and the linearized approach be 
self consistent, the Zerilli function $\Phi_1^+$ has to satisfy  
(\ref{series}) and (\ref{cond}) whenever
 $r_c<\ri$. It also has to satisfy condition (\ref{cond0}), and decay properly
as $r \to \ri$. 
In particular, both initial data functions 
 $\Phi_1^+(t=0,r), \dot \Phi_1^+(t=0,r)$ must satisfy all these conditions. \\

The rather odd initial value problem that these conditions pose is in fact very similar to that of the propagation of
scalar gravitational perturbations on a negative mass Schwarzschild background, which also has
a ``kinematic'' singularity. This latter
problem was worked out in \cite{nm}  using an intertwiner operator \cite{price}. We will apply
the same technique in what follows.

\subsection{Basics of the intertwining technique} \label{basics}

Consider a wave equation with a time independent potential:
\begin{equation}  \label{s1}
  \frac{\partial \Phi}{\partial t^2} - \frac{\partial \Phi}
{\partial x^2} + V \Phi  =: \frac{\partial \Phi}{\partial t^2}
+ {\cal H} \Phi = 0
\end{equation}
on a domain $t \in {\mathbb R}$, $x \in (a,b)$ where $a=-\infty$ and/or $b=\infty$ is a possibility.
 An intertwinner for this equation has the form \cite{nm,price}
\begin{equation} \label{inter1}
\I = \frac{\p}{\p x} - g, \hspace{1cm}  g= \frac{1}{\psi_{\cal I}} \; \frac{d \psi_{\cal I}}{d x}
\end{equation}
with $\psi_{\cal I}$ satisfying
\begin{equation} \label{ip}
{\cal H} \psi_{\cal I} = E_{{\cal I}} \psi_{\cal I}.
\end{equation}
The above equations neither  assume that ${\cal H}$ is self adjoint in $L^2((a,b),dx)$ nor that
$\psi_{\cal I}$ is an eigenfunction of such an operator. $\psi_{\cal I}$ in equation (\ref{ip}) is
just {\em any} solution of this differential equation, without any consideration on boundary conditions,
boundedness or finiteness of  some $L^2$ norm.\\
In terms of
\begin{equation} \label{hf}
\hat \Phi := \I \Phi
\end{equation}
 equation (\ref{s1}) reads
\begin{eqnarray} \nonumber
  0 &=& \frac{\partial \hat \Phi}{\partial t^2} + \hat {\cal H} \hat \Phi \\
\hat {\cal H} &:=& -\frac{\partial \hat \Phi}{\partial x^2} + \hat V \label{hz} \\
 \hat V &:=& V - 2 \frac{d g}{d x} \nonumber
\end{eqnarray}
That is, if $\Phi$ is a solution of (\ref{s1})
with initial data
\begin{equation}\label{id}
(\Phi(t=0,x), \dot \Phi(t=0,x)),
\end{equation}
 then $\I \Phi =: \hat \Phi$
is a solution of (\ref{hz}) with initial data
\begin{equation} \label{hid}
(\hat \Phi(t=0,x), \dot {\hat \Phi}(t=0,x))=(\I \Phi(t=0,x), \I \dot \Phi(t=0,x)).
\end{equation}
The general idea of the intertwiner technique is to use this fact to search for  an appropriate
intertwiner such that  $\hat V$ is simpler than the potential in the original problem. \\
The operator $\I$ has a nontrivial kernel spanned by $\psi_{\cal I}$, so information is lost when switching from
$\Phi$ to $\hat \Phi := \I \Phi$. \\
The operator
\begin{equation} \label{inter}
\hat \I = \frac{\p}{\p x} + g,
\end{equation}
can easily be seen to map solutions of (\ref{hz}) onto solutions of (\ref{s1}).
A straightforward computation shows that
\begin{equation} \label{ii}
\hat \I \, \I = E_{{\cal I}}- {\cal H}
\end{equation}
Since $\I$ and $\hat \I$ have non trivial kernels, information is lost when applying these
operators.
However, in the case of a solution of the wave equation (\ref{ii}) implies that
\begin{equation}
\hat \I \hat \Phi = \hat \I \, \I  \Phi = ( E_{{\cal I}}- {\cal H}) \Phi =  E_{{\cal I}} \Phi +\frac{\p^2  \Phi}{\p t^2},
\end{equation}
 Thus the information lost is precisely the $\Phi$ initial data. The above equation
can be regarded as a $t-$ODE on $\hat \Phi$ for every $x$, and can easily be integrated
to give $\Phi$ back:\\
If $E_{\I} >0$,
\begin{equation} \label{conv1}
\Phi(t,x) = \frac{1}{\sqrt{E_{\I}}} \int_0^t \sin \left(\sqrt{E_{\I}} (t-t') \right) \; \hat \I \hat \Phi(t',x) dt' +
\cos ( \sqrt{E_{\I}} t ) \Phi(0,x) + \frac{\sin ( \sqrt{E_{\I}} t )}{\sqrt{E_{\I}}} \dot \Phi(0,x);
\end{equation}
if $E_{\I}<0$,
\begin{equation} \label{conv2}
\Phi(t,x) = \frac{1}{\sqrt{-E_{\I}}} \int_0^t \sinh \left(\sqrt{-E_{\I}} (t-t') \right) \; \hat \I \hat \Phi(t',x) dt' +
\cosh ( \sqrt{-E_{\I}} t ) \Phi(0,x) + \frac{\sinh ( \sqrt{-E_{\I}} t )}{\sqrt{-E_{\I}}} \dot \Phi(0,x);
\end{equation}
if $E_{\I}=0$,
\begin{equation} \label{conv3}
\Phi(t,x) =  \int_0^t (t-t')  \; \hat \I \hat \Phi(t',x) dt' +
 \Phi(0,x) + t\; \dot \Phi(0,x) = \int_0^t \left( \int_0^{t'} \hat \I \hat \Phi(t'',x) dt'' \right) dt' +
 t \dot \Phi(0,x)
+ \Phi(0,x).
\end{equation}

We conclude that we can solve equation (\ref{s1}) subject to (\ref{id}) by means of
the following procedure:
\begin{enumerate}
\item  From the initial $\Phi$ data (\ref{id}) construct initial $\hat \Phi$ data using (\ref{hid}).
\item Find the solution $\hat \Phi$ of equation  (\ref{hz}) with initial data (\ref{hid}).
\item Apply  (\ref{conv1})-(\ref{conv3}) to obtain the  solution $\Phi$ to the original equation.
\end{enumerate}
Note that the {\em evolution problem} is solved in step 2, and that the initial $\Phi$ data
is used twice: in steps 1 and 3.\\

\subsection{The initial value problem for perturbations of region III}

Why would be one be interested in solving (\ref{s1})  using the complicated intertwiner method?  The intertwiner is certainly useful  if  $\hat V$ is  simpler than $V$.
Our motivation, however,  comes from a deeper problem: there is no available theory to deal with the initial
value problem possed in Lemma 1. Unless $r_c>\ri$, which may only happen for
a finite number of harmonic numbers $\ell$, the function space in Lemma 1 is unrelated
to any recognizable Hilbert space, and  the Zerilli wave equation has singular coefficients in region III
(on top of this, there is the issue of non global hyperbolicity of region III, even
when $r_c>\ri$.)
A similar situation is found when studying  perturbations of a negative mass Schwarzschild spacetime.
In this case, the problem was  solved using intertwiners \cite{nm}.
 Less sophisticated approaches, such as using algebraic redefinitions
of the variables, can be shown to fail.\\
 As we show in Appendix A, for {\em any} (even complex) $E_{\I}$, the
intertwiner (\ref{inter1})-(\ref{ip}) produces a $\hat V$ that is smooth at $r=r_c$, while sending
functions satisfying (\ref{series}) and (\ref{cond}) onto functions which are smooth at $r=r_c$.
 Also, there are
options for $\psi_{\I}$ for which $\I$ sends initial data (as characterized in Lemma 1) onto
a subspace ${\cal D} \subset L^2((0,\infty), dx)$ where $\hat {\cal H}$ is a self-adjoint operator.
This allows us to use the resolution of the identity for $\hat {\cal H}$ to solve the hat wave equation
by separation of variables, by expanding the initial data
using normalized eigenfunctions of $\hat {\cal H}$,  $\hat {\cal H} \hat \psi_E = E \hat \psi_E$:
\begin{eqnarray} \label{evo}
\hat \Phi(t,x) &=& \sum_E a_E(t,x) \hat \psi_E(x)\\ \label{evo2}
\hat \Phi(0,x) &=& \sum_E a_E^0 \hat \psi_E(x)\\
\dot{\hat \Phi}(t,x) &=& \sum_E \dot{a}_E^0 \hat \psi_E(x). \label{evo3}
\end{eqnarray}
Here the coefficients are obtained by integrating against the complex conjugate of $\hat \psi_E(x)$, and
the wave equation then reduces to an infinite set of ODEs:
\begin{eqnarray}\nonumber
 \ddot a_E &=& -E a_E \\  \label{b1}
 \dot a_E(0) &=& \dot a_E^0 := \int \overline{\hat \psi_E } \;\;\dot{\hat \Phi}(t=0,x) \; dx  \\
 a_E(0) &=& a_E^0 := \int \overline{\hat \psi_E}\;\; \hat \Phi(t=0,x) \; dx .  \nonumber
\end{eqnarray}
whose solution is
\begin{equation} \label{b3}
a_E(t) = \begin{cases} a_E^0 \; \cos (\sqrt{E} t) + \dot a_E^0 \; E^{-1/2} \; \sin(\sqrt{E} t) \ & , E>0\\
a_E^0 + t \; \dot a_E^0 &, E=0 \\
a_E^0 \; \cosh (\sqrt{-E} t) + \dot a_E^0 \; (-E)^{-1/2}\; \sinh(\sqrt{-E} t) & , E<0
\end{cases}
\end{equation}

The above equations {\em define} the evolution of the fields outside the domain of dependence
of the initial data. This same technique was applied, e.g., in  \cite{wi,nm}, in similar contexts.
A subtle issue is that of defining
the domain ${\cal D} \subset L^2((0,\infty),dx)$
where $\hat {\cal H}$ is self-adjoint. This problem
is identical to that of quantum mechanics on a half axis,  treated
in detail in the first reference in \cite{nc}. Consider the two dimensional
vector space of local (Frobenius)
solutions of the eigenvalue equation $\hat {\cal H} \hat \psi = E \hat \psi,
\hat \psi \neq 0$, near $x=0$. Given that an overall factor on $\hat \psi$ is irrelevant, the space of
local solutions can be regarded as a circle (this is why we used $A \cos(\theta)$ and
$A \sin (\theta)$ for the two arbitrary constants  in equations (\ref{asyms}),
(\ref{asymv}), etc. $\theta \in [0,2\pi)$ labels points in this circle.)\\
 If any eigenfunction is square integrable near $x=0$ we
say, following \cite{nc}, that $\hat {\cal H}$ belongs to the ``limit circle case''.  In this case,
$\hat {\cal H}$ will be a self-adjoint operator only after restricting to a  subspace
${\cal D}_{\theta_o} \subset L^2((0,\infty),dx)$ of functions behaving near $x=0$ as local eigenfunctions with
a fixed $\theta=\theta_o$. Equations (\ref{evo})-(\ref{b3}) will then hold in  ${\cal D}_{\theta_o}$, for initial data
in this space. Note, however, that initial data of compact support belongs to ${\cal D}_{\theta}$ for {\em any} $\theta$,
and evolve in a  different way if some $\theta_o' \neq \theta_o$ is chosen in (\ref{evo})-(\ref{b3}).
Of course, the solution
will be different  {\em only outside the domain of
dependence of the initial data}, but still there is an ambiguity, which must be resolved.
Physical input must then dictate what the right choice of $\theta$ is
in order to get rid of this ambiguity.\\
The other possibility is that the hamiltonian  piece of the wave equation
 belongs to the ``limit point case'', i.e., that there is
a single $\theta$ value giving local solutions which are square integrable near $x=0$. In this case
we say that $\hat {\cal H}$ is ``essentially self-adjoint'' (since it is only self-adjoint
in the domain defined by this particular $\theta$ value) and there is no ambiguity in the dynamics.
This would be the case if one of the roots of the indicial equation of of the Frobenius local solution of
the hamiltonian eigenfunction were less than $-1/2$.\\
For the scalar gravitational perturbation problem, the situation is that of a limit circle hamiltonian.
 The self consistency condition (\ref{cond0}), however, singles out
 a unique  $\theta$. With this choice,
the degree of divergency as $x \to 0^+$ not only of $R_1$, but also of   {\em all} the remaining
algebraic invariants of the Riemann tensor, gets controlled, and,
since the evolution (\ref{evo})-(\ref{b3})  preserves the local behaviour at $x=0$,
the invariants will stay properly bounded near the singularity at later times.
The dynamics is thus unambiguous once we enforce the self consistency condition (\ref{cond0}).\\

In \cite{nm}, a ``zero mode'' ($E_{\I}=0$) was used to construct the intertwiner and produce
a self-adjoint $\hat {\cal H}$. The resulting hamiltonian has a negative energy eigenvalue, and thus
exhibits the instability of the spacetime. The Zerilli field is then recovered
using equation (\ref{conv2}), from where it is clear that the exponential growing in time  of $\hat \Phi$ shows up
in the metric and electromagnetic field perturbations.
We have tried this same approach here, and found that
 an appropriate zero mode can be explicitly constructed for $\ell=2$ and, as happens
in the Schwarzschild naked singularity
case. $\hat {\cal H}$ has a smooth potential and contains a negative energy eigenvalue,
at least for some $Q/M$ values for which $r_c<\ri$ (see  Section \ref{l=2}).
Since generic  perturbation initial data  with compact support in region III
will have a nonzero projection onto the $\ell=2$ type-one scalar mode, this is certainly
enough to  show that such initial data will excite unstable mode in these cases.\\
However, we were not able to prove that for
{\em arbitrary}  $\ell$ and $Q/M$ such that $r_c < \ri$ there is a  zero mode  intertwiner which
does not introduce a {\em new} singularity in $\hat V$
 (although it is still trivial to show that any intertwiner
washes out the singularity at $r=r_c$, see Appendix A.) A new singularity would be introduced if $\psi_{\I}$ had
a zero for $r \in (0,r_c) \cup (r_c,\ri)$. \\

For this reason,  in Section \ref{chandrain}  we exhibit an alternative intertwiner for which computations
can be carried out explicitly for any $\ell$. This uses $\psi_I = \chi_1^+$, Chandrasekhar's mode,
and gives, for any $Q/M$ and $\ell$,
 $\hat {\cal H} = {\cal H}^-_1$ (the Hamiltonian for type-1 {\em vector perturbations}!).
Since $\hat {\cal H}$ is positive definite in this case, the hat wave equation is stable, and the scalar instability
shows up only when reconstructing the Zerilli field using (\ref{conv2}). This is so because
those factors inside the integral
which are exponential in
$t$ do not cancel the exponential factors outside the integral (as it would happen if the original wave equation were stable).
These factors will then show up in the metric and electromagnetic field perturbations, the Riemann tensor, and
its invariants.

\subsection{The $\ell=2$  zero mode  intertwiner}
\label{l=2}

In \cite{nm}, a ``zero mode'' (solution of ${\cal H} \psi=0$ that is not necessarily
normalizable or well behaved at the boundaries) was used to construct an
intertwiner to deal with the initial value problem for the scalar mode negative mass Schwarzschild perturbations,
which has difficulties similar to those found in the present case. \\
We may try the same approach here, however, given
the complexity of the potential in  ${\cal H}^+_1$, it is rather difficult to obtain
the solutions of
\begin{equation} \label{szm}
{\cal H}^+_1  \psi_{o}^+ =0
\end{equation}
for $\psi_{\I} = \psi_o^+$
required to construct the $E_{\I}=0$ intertwiner $\I$  (\ref{inter1}).
One possibility is to use
the relations
 (\ref{fsh})-(\ref{fvh}) to obtain scalar zero modes from vector ones, since,
for $ \psi_o^-$ a {\em vector} zero mode,
\begin{equation} \label{vzm}
{\cal H}^-_1 \psi_o^- = 0,
\end{equation}
it follows from  (\ref{fsh})-(\ref{fvh}) that
$ {\cal A}_1 \psi^-_o$ is a {\em scalar} zero mode. Since the vector potential $V^-_1$ is much
simpler than  $V^+_1$, there is some hope that we could  carry on calculations in a more explicit way
using this idea. This is indeed the case for  $\ell=2$, for which
the general solution of equation (\ref{vzm}) (see Appendix B) can be shown to be
\begin{multline} \label{l2zvm}
\psi_o^-= A \cos(\a) \;\left\{ {r}^{3}+ \frac{\beta_2}{4} (Q^2-r^2)
-{\frac {{Q}^{4}}{r}} \right\}\\
 + A \sin (\a)  \left\{
3\, \left( 4\,{r}-\beta_2  +4\,\frac{{Q}^{2}}{r}
 \right)
\left( \frac{{r}^{2}-{Q}^{2}}{2 \sqrt{{M}^{2}-{Q}^{2}}} \right)  \left[ \ln  \left(
\frac{-r+M}{\sqrt {{M}^{2}-{Q}^{2}}}-1 \right) -\ln  \left(
\frac {-r+M}{\sqrt {{M}^{2}-{Q}^{2}}}+1 \right)  \right]   \right. \\ \left.
 -\left( 12\,{r}^{2}+3\, \left(\beta_1 - 2M \right) r+  \left(
 3M \beta_1 -2 (M^2 + 2 Q^2) \right) + \frac{12\,{M}^{3}-2\,
 (\beta_2 - \beta_1)  \left( {M}^{2}-{Q}^{2} \right) }{r} \right)
 \right\} .
\end{multline}
where $A$ and $\a$ are arbitrary constants. Note that
the $\ell=2$ intertwiner operator constructed using $\psi_{\I} = \psi_o^+ = {\cal A}_1 \psi^-_o$ in (\ref{inter1})
will depend on $\a$ but certainly not on $A$. It can be easily shown, however, that
 $\hat V$ is smooth at $r=r_c$ irrespective of the choice of $\a$,
 the first and second order poles in $V_1^+$ being
canceled by the poles in $dg/dx$ (Figure 4). This, of course, is to be expected from the
more general considerations in Appendix A. The asymptotic behaviour of $\hat V$
near the singularity and the inner horizon is also independent of $\a$:
\begin{equation}
\hat V \simeq \begin{cases} \frac{4}{9 x{}^2} + ...  & , x \simeq 0\\
        C \exp \left(
 -\frac{(\ro-\ri)x}{\ri^2} \right) & , x \to \infty \end{cases}
\end{equation}
Near the singularity
the eigenfunctions of $\hat {\cal H}$ behave as
\begin{equation} \label{hzm}
\hat \psi =  A  \cos(\theta)  [ x{}^{-1/3} \;  \sum_{n \geq 0} a_n^{(1)} x^{n/3} ] +
 A \; \sin(\theta) [ x{}^{4/3}  \sum_{n \geq 0} a_n^{(2)} x^{n/3} ]
\end{equation}
Thus $\hat {\cal H}$ belongs to the limit circle case,
and only restricting to a subspace ${\cal D}_{\theta_o}$ of functions behaving as (\ref{hzm}) with a fixed $\theta
=\theta_o$ value
 does $\hat {\cal H}$ becomes self-adjoint. \\
We will make the choice $\theta=\pi/2$ of slowest decaying functions. This condition is certainly preserved by
the hat wave equation (see (\ref{evo})), and implies
\begin{equation} \label{ci}
\hat \I \hat \Phi = \sum_{k \geq 1} a_k r^k, \;\;\; a_2 = \frac{\b_2}{4Q^2} \, a_1,
\end{equation}
near $r=0$. If $\Phi_1^+(0,r), \dot \Phi_1^+(0,r)$ also admit expansions like those in (\ref{ci}),
then, using  (\ref{conv2}) follows that
\begin{equation} \label{cig}
 \Phi_1^+(t,r) =  \sum_{k \geq 1} a_k(t) r^k, \;\;\; a_2(t) = \frac{\b_2}{4Q^2} \, a_1(t),
\end{equation}
for all $t$,
 a condition that can be shown to guarantee that
{\em all} algebraic invariants of the Riemann tensor behave properly near the singularity.\\

Regarding the choice of $\alpha$ in (\ref{l2zvm}), although the results do not depend on the intertwiner
that we use, it is certainly easier to understand how the instability is excited by an initially compactly bounded
perturbation if we use
\begin{equation} \label{l2zvm2}
\tan (\a) = \frac{2Q^2 \, \beta_2 \, \sqrt{M^2-Q^2}}{3 Q^2 \beta_2 \, \ln \left( \frac{M-\sqrt{M^2-Q^2}}{M+ \sqrt{M^2+Q^2}}
\right) + 2 ( 3M \beta_2-16 (M^2 - Q^2)) \sqrt{M^2-Q^2}} ,
\end{equation}
since in this case
the resulting intertwiner will send $\chi_1^+$ onto the the Hilbert space ${\cal D}_{\pi/2}$ selected
by the self consistency argument, and thus $\I \chi_1^+$ will be one of the eigenfunctions of $\hat {\cal H}$
(it will actually be the only negative energy  $\hat {\cal H}$ eigenfunction).
The transformed potential $\hat V$, together with  $\I \chi_1^+$ for this choice are given in Figure 4 below for
some specific $Q$ and $M$ values.
An explicit expression for $\hat V$ can be readily obtained using equations (\ref{inter1}), (\ref{hz}),
$\psi_o= {\cal A}_1 \psi_o^-$, (\ref{l2zvm}) and  (\ref{l2zvm2}). \\
Now suppose some perturbation data $(\Phi_1^+(t=0,x), \dot \Phi_1^+(t=0,x))$ of compact support is given.
The hat wave equation data $(\I \Phi_1^+(t=0,x), \I \dot \Phi_1^+(t=0,x))$ will be of compact support and
then it will
belong to ${\cal D}_{\pi/2}$. Expanding it using (\ref{evo2})-(\ref{evo3})  will generically give a nonzero projection onto
the fundamental, unstable  $\hat {\cal H}$ eigenfunction  $\I \chi_1^+$, and thus, from (\ref{b3}),  an exponentially
growing term in (\ref{evo}), which survives when $\Phi^+_1$ is reconstructed using (\ref{conv3}) and shows up in the
metric and electromagnetic field perturbations.\\

The use of a zero mode intertwiner has some drawbacks: although we can show that the kinematic singularity
is absent from $\hat V$  (see Appendix B), we do not have
a complete proof, even for $\ell=2$, that the 
zero mode $\psi_{\I}=\psi_o^+$ has no zeroes in $(0,r_c) \cup (r_c,\ri)$, which would 
introduce new singularities in $\hat V$.
However, for $\ell=2$, we have numerically verified that this is the case for a range of values of $Q/M$. 
A particular example of a smooth $\hat V$ for $\ell=2$ is  that given in Figure 4.\\

In the following Section, we show that all these issues can be avoided by using an alternative intertwiner that
allows explicit calculations for every harmonic number and charge and mass values.

\begin{figure}[h]
\centerline{\includegraphics[width=9cm,height=6cm]{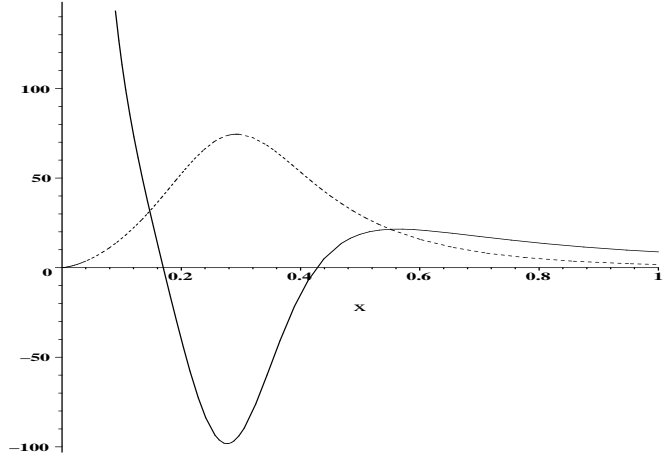}}
\caption{$\ell=2$ potential $\hat V$ -continuous line-
for the transformed Zerilli equation (\ref{hz}), and the transformed of the unstable mode $\chi_1^+$.
In this example $r_i=1$ and
$r_o=2$.}
\end{figure}

\subsection{Intertwining using Chandrasekhar's algebraic special mode} \label{chandrain}

We can apply the intertwining technique using Chandrasekhar's algebraic special mode $\chi_1^{+}$ in
(\ref{inter1}), for which
\begin{equation}
E_{\I} = - \left(\frac{\kappa}{2 \beta_1} \right)^2.
\end{equation}
Using this intertwiner has a number of advantages.
 Contrary to what happens for the zero mode, we have an explicit expression for
$\chi_1^+$ for every harmonic number $\ell$, equation (\ref{chi}).
 It is also simple to construct $\hat V$ using
(\ref{pots}) and (\ref{chis}) (a prime denotes derivative with respect to $x$):
\begin{equation} \label{ppq}
\hat V = V_1^+ - 2 \left( \frac{{\chi_1^+}'}{\chi_1^+} \right)' =
V_1^+ - \left( \beta_{1} f_{1} + \frac{\kappa}{2 \beta_{1}} \right)' = V_1^- =
\frac{f}{r^4} \left( \ell(\ell+1) r^2- \beta_1 r + 4 Q^2 \right)
\end{equation}
(This relation between the scalar and vector modes was first noticed by Chandrasekhar \cite{prs,chandra}.)
The fact that $\hat V = V_1^-$, the type-1 vector potential, simplifies the analysis considerably,
since it is clear that  $V_1^-$ is smooth in $(0,\ri)$ for {\em any} value of $Q/M$.\\

From the first line in (\ref{asymv}) follows  that $\hat {\cal H} = {\cal H}^-_1$ belongs
to the limit circle case. As explained above, a choice  $\theta_o$ has to be made to fix
the domain ${\cal D}_{\theta_o} \subset L^2((0,\infty), dx)$ where $\hat {\cal H}$ is self adjoint.
However, for type-1 scalar perturbation the consistency requirement (\ref{cond0}) (see also Lemma 1) reduces to:
\begin{equation}
 \Big | \I \Phi_1^+ \Big | \leq N,  \;\;
\text{ if } \;\; 0 < r < \d
\end{equation}
which rules the $x^{-1/3}$ piece of (\ref{asymv}), and forces $\theta=\pi/2$.
Once again,
this condition is  preserved by
the hat wave equation (see (\ref{evo})), and imply (\ref{ci}) and (\ref{cig}) for initial data
satisfying this condition (where now $\I$ has to be understood as the intertwiner made
using Chandrasekhar's mode)
Condition (\ref{cig})  guarantees that
{\em all} algebraic invariants of the Riemann tensor behave properly near the singularity at later times.\\

Since
$\hat V$ (and thus $\hat {\cal H}$) is positive definite, the hat wave equation is stable, and its modes
oscillate in time. The instability of $\Phi_1^+$ arises as a consequence of the fact that, generically,
 the exponential terms
in the integrand of (\ref{conv2}) do not cancel out with those outside the integral.
 As a trivial example, take $\Phi_1^+(t=0,x)$ and
$\dot \Phi_1^+(t=0,x)$ both proportional to $\chi_1^+$ (this certainly passes all the requirements in Lemma 1).
 Since $\I \chi_1^+ =0$, the initial data in hat space is trivial, then $\hat \Phi (t,x)=0$ for all $t$,
and (\ref{conv2}) reduces to the last two terms, which are generically exponentially growing  for large $t$.\\

A final observation is the nontrivial fact that, unlike the Zerilli field,
 the intertwined variable has a geometrical significance, as, from (\ref{dr1}),  it gives the first order variation
of the Riemann invariant $R_1$:
\begin{equation} \label{rel}
\I \Phi^+_1 = \hat \Phi = - \frac{r^8}{4 Q^2} \;  \d R_1
\end{equation}
This gives further significance to the dual relation between vector and scalar modes first found by Chandrasekhar,
which was limited to some observations on their mode spectra. Equations (\ref{dr1}) and (\ref{ppq}) prove that the field 
giving the first order 
variation of $R_1$ (times $Q^4/r^8$) associated to a {\em scalar} mode perturbation is a solution of
the {\em vector} mode perturbation equation of the same harmonic number.

\section{Conclusions}

We have proved that the inner static region $0 < r < \ri$ of a \rn black hole
is unstable under linear perturbations of the metric and electromagnetic field.
More precisely, we
have shown that a perturbation with compact support within this region
will excite unstable type-1 polar modes, of which there is one for every harmonic number
$(\ell,m)$. This instability is relevant to the strong cosmic censorship
conjecture, according to which this region of the black hole, lying beyond the Cauchy
horizon (of a Cauchy surface like the one in Figure 1), should be disregarded, as
it could not arise as a result of the collapse of ordinary matter departing
from spherical symmetry.\\
This result has implications on some simple models of halted collapse of a pressure-less charged perfect fluid star
\cite{smpc}, according to which the world tube of the surface of the star traces a path going
from the (right copy of) region I in Figure 1, through region II, into the {\it left} copy
of region III, upper copy of region II, then upper right copy of region I. To the right of this curve,
this spacetime
agrees with the extended \rn spacetime,
 which contains an entire copy of region III, thus being  unstable.
 We are currently studying these models in more detail.\\
The difficulty in establishing in a rigorous way the \rn instability lies in the fact that
the field variable which succeeds in disentangling the linearized equations, the Zerilli
field $\Phi_1^+$, happens to be a singular function of the perturbed fields in region III. This
cannot be cured by any simple field redefinition, but requires the use of an intertwiner operator
${\cal I} = \p / \p x + g$ that maps to a smooth  field $\hat \Phi := {\cal I} \Phi_1^+$. The information
lost due to the nontrivial kernel of ${\cal I}$ is entirely contained in the initial data, and thus
is available. The evolution of perturbations on the non globally hyperbolic background is well
defined by using the spectral theorem and a unique self-adjoint extension of the spatial piece
of the wave operator that gives the dynamics of the $\hat \Phi$ field. We should comment here
that intertwiners in the context of linear perturbations were first considered in \cite{price},
while they were first used to deal with  the issue of the singularities of the Zerilli field
in the proof of the instability of the Schwarzschild naked singularity in \cite{nm}. The idea of
defining  dynamics on non globally hyperbolic backgrounds by using a suitable self adjoint extension
of the spatial piece of the wave equation together with the spectral theorem was first suggested
in \cite{wi}. The main difference between the cases considered in \cite{wi}, and the \rn
and negative mass Schwarzschild cases, lies in the fact that the spatial operators (``Hamiltonian'') in the last
two cases are not positive definite. A difference between the negative mass Schwarzschild
and the \rn cases, is that the intertwiner used in the first case gives a Hamiltonian
with a single self-adjoint extension (limit point case in \cite{nc}), whereas
the one for \rn corresponds to the limit circle case in \cite{nc}. \\
Two different intertwiners were used: one constructed out of a zero mode, the other using one of
Chandrasekhar's  algebraic special modes. The first intertwiner has the advantage of exhibiting the instability
in a rather obvious way, and the drawback that we lack explicit expressions for the interwtined potential,
or a proof of its smoothness within the relevant parameter range. The intertwiner that
uses Chandrasekhar's ``hides'' the instability, which is made explicit
in the metric reconstruction process through the original Zerilli field. This mode allows explicit calculations
for every harmonic number and $Q$ and $M$ values. It also exhibits a very interesting connection between
vector and scalar modes: the intertwined field $\hat \Phi$ gives the first order variation
of the Riemann invariant $R_1$ (see equation (\ref{rel}). An alternative way of stating this is that
the first order variation $\delta R_1$ of $R_1$ under {\em scalar} perturbations is a solution
of the Zerilli {\em vector} perturbation equation.\\
The results presented here
adapt easily to the case of a super-extreme ($|Q|>M$) \rn spacetime, and thus can be
used to fill in the details left untreated in \cite{doglepu} to show that a perturbation
of an overcharged \rn spacetime, compactly
supported away from the singularity, will excite modes that grow exponentially in time.
This, of course, is relevant to weak cosmic censorship, this being the original motivation
for our work.

\section*{Acknowledgements}

This work was supported by grants PIP 112-200801-02479 from CONICET (Argentina),
Secyt 05/B384 and 05/B253 from Universidad Nacional de C\'ordoba (Argentina), and a Partner Group grant from
 the Max Planck
Institute for Gravitational Physics, Albert-Einstein-
Institute (Germany).  RJG and GD are supported by
CONICET. GD wishes to thank the  AEI,
where part of this work was done, for hospitality and support.

\appendix

\section{Intertwined potential} \label{hl}

In this section we analyze the local behaviour of the intertwined potential and
 $\hat \psi$ given in equations (\ref{inter1}),(\ref{ip}) and  (\ref{hz}) as the singularity, inner
horizon, and $r=r_c$ are approached.
 This is done
by studying the  behaviour of generic solutions of the equation
\begin{equation}\label{app01}
{\cal H}_1^+ \psi_{\I} =
-f \frac{d}{dr}\left(f\frac{d \psi_{\I}}{dr}\right) +V^+_1 \psi_{\I} = E_{\I} \psi^+_1
\end{equation}
for $r \simeq 0, r \simeq r_c$, and $r \simeq \ri$. \\

Note that  $V^+_1$ may be written as,
\begin{equation}\label{app02}
V^+_1 = \frac{f}{r \left(\beta_1+(\ell-1)(\ell+2)r\right)^2}\;
 \left[\left(\kappa+\beta_1\dfrac{df}{dr}\right)\left(\beta_1+(\ell-1)(\ell+2)r\right) -
2 f(r)\beta_1(\ell-1)(\ell+2)\right]
\end{equation}
showing explicitly the double pole at,
\begin{equation}\label{app03}
r = r_c = -\frac{\beta_1}{(\ell-1)(\ell+2)}.
\end{equation}

\subsection{Behaviour of $\hat V$}
Near $r=r_c$, the general solution of (\ref{app01}) is of the form,
\begin{multline}
\psi_{\I} = \frac{1}{ r -r_c}\left[ c_0  - \frac{(\ell-1)^2(\ell+2)^2(2M-\beta_1) \; c_0 }{\beta_1\left(2(\ell-1)
(\ell+2)M +(\ell^2+\ell+2)\beta_1\right)}(r-r_c)\right.   \\
  \left. + \frac{8 E_{\I} \beta_1^2 \; c_0}{ \left(2(\ell-1)(\ell+2)M +(\ell^2+\ell+2)\beta_1\right)^2}(r-r_c)^2
+ c_3 (r-r_c)^3 + c_4 (r-r_c)^4 + \dots\right] \label{app05}
\end{multline}
where $c_0$ and $c_3$ are arbitrary constants, and $c_4$ and higher coefficients
in the series are linear combinations of  $c_0$ and  $c_3$ with coefficients that depend on $E_{\I}, Q, M$ and $\ell$.
Note that the generic local eigenfunction above satisfies the requirement (\ref{cond}).\\

If we use the generic  $\psi_{\I}$ given above to construct the   potential $\widehat{V}$,
\begin{equation}\label{app06}
\widehat{V}= V - 2 f\frac{d}{dr}\left(\frac{f}{\psi_{\I}}\frac{d}{dr}\psi_{\I}\right)
\end{equation}
a straightforward computation shows that, provided $a_0 \neq 0$, near $r=r_c$,
\begin{equation}\label{app07}
\widehat{V}= \frac{8 k^2 \beta_1^4 +(\ell+2)^3(\ell-1)^3[(\ell^2+\ell+4)
\beta_1^2-20 M \beta_1 -12(\ell+2)(\ell-1)M^2]}{4 \beta_1^4} + {\cal{O}}(r-r_c).
\end{equation}
This means that, for {\em any} $E_{\I}$, an {\em arbitrary} solution of (\ref{app01}) with $a_0 \neq 0$ gives
an intertwined potential $\widehat{V}$ that is smooth at $r=r_c$.
We notice also that, as can be checked, if $a_0=0$, the second term in the R.H.S. of (\ref{app06}) does not compensate the double
pole in $V$, so that $\widehat{V}$ is also singular.\\

We consider next the local behaviour  near $r=0$.
The general solution of (\ref{app01}) admits an expansion in powers of $r$ of the form,
\begin{equation}\label{app08}
\psi_{\I}(r) = a_1 r +a_2 r^2 + \left[ \frac{(\ell-1)(\ell+2)((\ell-1)(\ell+2)Q^2+M\beta_1)}{Q^2\beta_1^2}a_1
 +\frac{M}{Q^2} a_2\right]r^3 + a_4 r^4 + \dots
\end{equation}
where $a_1$, and $a_2$ are arbitrary constants and the higher order coefficients depend linearly on them.
A dependence on $E_{\I}$ appears first at order $r^7$.
This result implies that, near $r=0$, assuming $a_1\neq0$, we have,
\begin{equation}\label{app10}
\widehat{V}=   \frac{4}{9} x{}^{-2}- \frac{2}{ 3^{5/3}} \left(\frac{M}{Q^{4/3}}  - \frac{2a_2 Q^{2/3}}{a_1}  \right)
 x{}^{-5/3} + {\cal{O}}(x{}^{-4/3})
 \end{equation}
while, if $a_1=0$,
\begin{equation}
\label{app10b}
\widehat{V}=\frac{10}{9} x{}^{-2} + \frac{3^{2/3}}{108} \left(\frac{10 \ell (\ell +1)-12)}{Q^{2/3}}+\frac{M(28 M - 5 \beta_1)}{Q^{8/3}}
\right)
 x{}^{-4/3} +
{\cal{O}}(x{}^{-1})
 \end{equation}

Finally we consider the behaviour near $r=r_i$. The cases $E_{\I} \neq 0$ and $E_{\I}=0$ require separate treatment.
For $E_{\I} \neq 0$, since the potential vanishes for $r=r_i$, the leading terms
of the two linearly independent parts of the solution for real $E_{\I}$ are of the form,
\begin{eqnarray}
\psi_{\I} & = & C_1 (r_i-r)^{\left(\dfrac{r_i^2 \sqrt{E_{\I}}}{r_o-r_i}\right)}+
 C_2 (r_i-r)^{\left(\dfrac{-r_i^2 \sqrt{E_{\I}}}{r_o-r_i}\right)} \nonumber \\
  & = & \tilde{C}_1 \exp \left( \sqrt{E_{\I}} x \right)+ \tilde{C}_2 \exp \left(-\sqrt{E_{\I}} x
\right)
\end{eqnarray}
where $C_1$, $C_2$, $\tilde{C}_1$, and $\tilde{C}_2$ are constants.
For $E_{\I} = 0$, on the other hand, the general solution admits an expansion of the form,
\begin{eqnarray}
\label{app21}
\psi_1^+(r) &=& a_0 + \frac{\left((r_i-r_o)\beta_1+\kappa r_i^2 \right) a_0 -  r_i\left(\beta_1+(\ell+2)(\ell-1)((2 \ell^2+2 \ell-1)r_i +2 r_o)\right) b_0}{r_i(r_i-r_o)(\beta_1+r_i(\ell+2)(\ell-1))}  (r-r_i) \nonumber \\
   & & + a_2 (r-r_i)^2+ \dots   \\
   & &  + \ln(r_i-r)\left[b_0 + \frac{\left((r_i-r_o)\beta_1 +\kappa r_i^2\right) b_0}{r_i(r_i-r_o)
(\beta_1+r_i(\ell+2)(\ell-1))} (r-r_i) + b_2 (r-r_i)^2+ \dots  \right].  \nonumber
\end{eqnarray}
where $a_2$, $b_2$ and higher order coefficients depend linearly on $a_0$ and $b_0$. \\
The $E_{\I}=0$  transformed potential behaves as
\begin{equation}\label{app22}
\widehat{V}(r)= \frac{2(r_o-r_i)^2b_0^2}{(a_0+b_0\ln(r_i-r))^2 r_i^4} + \dots
\end{equation}
where the dots indicate terms that vanish as a $(r_i-r)$. In terms of $x$, for $b_0 \neq 0$, this implies,
\begin{equation}\label{app22a}
\widehat{V}(x)= \frac{2 }{x^2} + \dots
\end{equation}

\subsection{Behaviour of $\hat \psi$}

We consider now the behaviour of the intertwined field $\widehat{\psi}$  (\ref{inter1}),
 which can be written as
\begin{equation}\label{inter2}
\widehat{\psi} = \I \psi = f  \psi_{\I}\frac{d}{dr}\left(\frac{\psi}{ \psi_{\I}}\right).
\end{equation}
We will use the following result, whose proof is straightforward:\\

\noindent
{\bf Lemma 2:} Assume that $\psi$ and $\psi_{\I}$  admit a Laurent expansion
\begin{equation}
\psi = (r-r_o)^p \; \sum_{k \geq 0} a_k \; (r-r_o)^k, \;\;\;\psi_{\I} = (r-r_o)^p \; \sum_{k \geq 0} a_k^{\I} \; (r-r_o)^k
\end{equation}
where $a_0=1=a_0^{\I}$,  and $p$ is any integer number. If $s$ is the highest number for which
$a_k = a_k^{\I}$ for every $ k \leq s$ ($s$ measures the degree of contact of these functions at $r_o$,
and, generically, $s=0$), then
\begin{equation}
\psi_{\I}\frac{d}{dr}\left(\frac{\psi}{ \psi_{\I}}\right) = (r-r_o)^p \; \sum_{k \geq s} d_k (r-r_o)^k, \;\; d_s = s (a_{s+1} - a_{s+1}^{\I}).
\end{equation}

Consider first the action of the intertwiner on a function $\psi$ satisfying (\ref{series}) and (\ref{cond}).
Since, as follows from (\ref{app05}), $\psi_I$ satisfies this same condition, generically $\psi$ and $\psi_{\I}$
have (as functions of $r$) degree of contact $s=2$, in the notation of Lemma 2, and thus $\hat \psi =
f \psi_{\I}\frac{d}{dr}\left(\frac{\psi}{ \psi_{\I}}\right)$ will be smooth at $r=r_c$.\\

The local solutions of ${\cal H}^+_1 \psi = E \psi$,
are of the form $\psi =  \sum_{k \geq 1} a_k r^k$ with $a_2$ and $a_1$ arbitrary, $a_k$ independent
of $E$ up to $k=7$.  Consider the action of $\I$ on a function like this further  subject to the condition
(\ref{ci}). If the intertwiner also satisfies (\ref{ci}), the degree of contact will be $s=5$, then
$ \psi_{\I}\frac{d}{dr}\left(\frac{\psi}{ \psi_{\I}}\right)$ will be ${\cal O} (r^6)$,  and thus $\hat \psi$
will be ${\cal O} (r^4)= {\cal O} (x^{4/3})$.

\section{The scalar and vector zero modes.}

In this Appendix
 we describe a procedure that allows the construction of the zero mode solutions for both vector and scalar modes. We start with the vector zero modes by first considering the differential equation they satisfy.
In accordance with (\ref{eqp}, \ref{pots}), and (\ref{eqp2}) with $\omega=0$, this is given by,
\begin{equation}\label{aw01}
-f^2 \frac{d^2\psi^{-}_0}{dr^2}-f \frac{df}{dr}\frac{d\psi^{-}_0}{dr}+ \frac{f}{r^4}\left( \ell (\ell+1) r^2-\beta_1 r +4 Q^2\right)\psi^{-}_0 = 0
\end{equation}

We notice that (\ref{aw01}) has regular singular points for $r=0$, $r=r_i$ and $r=r_o$, and no other singularity.
From now on we will write simply $\psi$ for $\psi_0^-$.

A simple analysis of the indicial equation shows that near $r=0$ this equation has two independent solutions,
one behaving as $r^{-1}$, and the other as $r^4$, both admitting a power series expansion.
We consider therefore an expansion for $\psi^{-}_0(r)$ of the form,
\begin{equation}\label{aw02}
\psi^{-}_0(r) = \frac{1}{r}\sum_{i=0}^{\infty}{a_i r^i}
\end{equation}
Replacing in (\ref{aw01}) we find that we must set $a_2=0$, and,
\begin{equation}\label{aw03}
a_1=-\frac{\beta_2}{4Q^2} a_0 \;\; ;\;\; a_3 =\frac{\ell(\ell+1)\beta_2}{24 Q^4} a_0\;\;;\;\; a_4= \frac{\ell(\ell^2-1)(\ell+2)}{24Q^4}a_0
\end{equation}
The coefficient $a_5$ can be chosen arbitrarily, in accordance with the indicial equation, and $a_6$ is given by,
\begin{equation}\label{aw04}
a_6 =-\frac{\ell(\ell^2-1)(\ell^2-4)(\ell+3)}{144 Q^6} a_0 -\frac{\beta_1-16M}{6 Q^2}a_5
\end{equation}

For the remaining coefficients, including $a_6$, we find a three term recursion relation of the form,
\begin{equation}\label{aw05}
-(\ell+j-2)(\ell-j+3) a_{j-1}+(\beta_1-2(j-1)(j-3)M)a_j+(j+1)(j-4)Q^2 a_{j+1}=0
\end{equation}
and, therefore, all the coefficients are determined once $a_0$, and $a_5$ are given. But this implies
 that for any given $\ell$, and $a_0 \neq 0$, we may choose $a_5$ such that $a_{\ell+3} =0$, and then
all coefficients for $j \geq \ell+3$ vanish. Calling $\psi_a$ this solution we have,
\begin{equation}\label{aw06}
\psi_a(r) = \frac{1}{r} {\cal{P}}_{\ell}(r)
\end{equation}
where ${\cal{P}}_{\ell}(r)$ is a polynomial of order $\ell+2$. The lowest order polynomials are,
\begin{eqnarray}
\label{aw07}
 {\cal{P}}_{2}(r)  &=& 1 -\frac{\beta_2}{4 Q^2} r +\frac{\beta_2}{4 Q^4}r^3 -\frac{1}{Q^4}  r^4 \nonumber \\
 {\cal{P}}_{3}(r)  &=& 1 -\frac{\beta_2}{4 Q^2} r +\frac{\beta_2}{2 Q^4}r^3 -\frac{5}{Q^4}  r^4 +\frac{30}{(\beta_2 +10 M) Q^4} r^5 \\
  {\cal{P}}_{4}(r)  &=& 1 -\frac{\beta_2}{4 Q^2} r +\frac{5 \beta_2}{6 Q^4}r^3 -\frac{15}{Q^4}  r^4 +\frac{21 (24M+\beta_2)}{4(3 Q^2 +M \beta_2 +6 M^2) Q^4} r^5 -\frac{42}{(3 Q^2 +6M^2 +M \beta_2)Q^4}r^6\nonumber \\
\end{eqnarray}
where we have fixed, for simplicity, ${\cal{P}}_{\ell}(r=0)=1$. With this normalization,
the polynomials are positive and decreasing functions of $r$ near $r=0$. We shall now prove that
they have no zeros in the interval $0 \leq r \leq r_i$. We write (\ref{aw01}) in the form,
\begin{equation}\label{aw08}
\frac{d^2\psi_a}{dr^2} = \frac{(2r_i r_o -(r_i+r_o)r)}{r(r_o-r)(r_i-r)}\frac{d\psi_a}{dr} +
\frac{(\ell(\ell+1)r^2-\beta_1 r +4 r_ir_o)}{r^2(r_o-r)(r_i-r)} \psi_a
\end{equation}
and notice that $\psi_a$ can have only simple zeros in $0 < r < r_i$
 because the coefficients in (\ref{aw08}) are regular functions of $r$ in $0 < r < r_i$.
Since sufficiently near $r=0$ we have $\psi_a >0$ and $d\psi_a/dr < 0$, and the
coefficients on the RHS in (\ref{aw08}) are both positive, the sign
 of $d^2\psi_a/dr^2$ is not fixed. We notice  however that at the first zero of $\psi_a$ for $r>0$ we must have $d\psi_a/dr < 0$,
 and therefore, we also have $d^2\psi_a/dr^2 < 0$. Since  to the right of such a  zero, and as long as $r< r_i$,
(since the coefficients are still positive) we must have both $d\psi_a/dr < 0$, and  $d^2\psi_a/dr^2 < 0$, and therefore
$\psi_a <0$, there can be no other zero for $r< r_i$. This proves that there is at most one zero for $r \in (0,\ri)$.\\
But now we notice that near $r=r_i$, equation (\ref{aw08}) has a singular
 solution (diverging as $\ln(r_i-r)$) and a unique regular solution  of the form,
\begin{equation}\label{ae09}
\psi(r) = \psi(r_i) \left[1 - \frac {(4 r_o+\ell(\ell+1)r_i-\beta_1)}{ r_i(r_o-r_i)} (r-r_i) + {\cal{O}}(r-r_i)^2 \right]
\end{equation}
Since $\psi_a$ is regular, it has the form (\ref{ae09}), and
 this implies that close to $r=r_i$ the regular solution and its first derivative have
 opposite signs. This contradicts the result obtained under the assumption that there is a zero in $0 < r < r_i$.
We conclude that $\psi_a$ does not vanish for $r \in (0,\ri)$.\\
Similarly, we find that near $r=r_o$, equation (\ref{aw08}) has a singular solution (diverging as $\ln(r_o-r)$) and
a unique non vanishing regular solution. This implies that  ${\cal{P}}_{\ell}(r)$ cannot vanish for $r=r_o$.

We note in passing that for the extreme case $Q=M$, where $r_i=r_o$, we have the exact (regular at the horizon) solutions,
\begin{equation}\label{aw10}
\psi_a(r)= \frac{C (\ell r +2M)(r-M)^{(\ell+1)}}{r}
\end{equation}
where $C$ is a constant.\\
Going back to (\ref{aw01}), for any fixed $\ell$, given the solution (\ref{aw06}), a linearly independent solution is given by,
\begin{equation}\label{aw11}
\psi_b(r)= C \frac{1}{r} {\cal{P}}_{\ell}(r) \int_0^r \frac{ y^4}{(y^2-2My+Q^2)\left({\cal{P}}_{\ell}(y)\right)^2} dy
\end{equation}
where $C$ is a constant. It is easy to check that $\psi_b$ is regular and non vanishing in $ 0 < r <r_i$, and,
\begin{eqnarray}
\label{aw12}
  \psi_b(r) &\sim & C_1 r^4 \;\;;\;\; r \to 0^+ \nonumber \\
  \psi_b(r) & \sim &  C_2 \ln (r_i-r) \;\;;\;\; r \to r_i{}^-
\end{eqnarray}
where $C_1$ and $C_2$ are constants. We may obtain and expansion of this solution in powers of $r$ using (\ref{aw11}), or directly from (\ref{aw01}),
\begin{eqnarray}\label{aw13}
\psi_b(r) & = & r^4 + \frac{(16M-\beta_1)r^5}{6 Q^2} + \frac{5 ((\ell^2+\ell-8) Q^2+4M(12M-\beta_1))r^6}{42 Q^4} \nonumber \\
& & +\frac{((\ell(\ell+1)(36M-\beta_1) +15 \beta_1-360M)Q^2+30 M^2 (32M-3 \beta_1))r^7}{84 Q^6} + \dots
\end{eqnarray}
where we have set an arbitrary multiplicative constant so that the coefficient of $r^4$ is equal to one.

For $\ell >2$ the integrals in (\ref{aw11}) cannot be computed directly, because that would require explicit expressions for the zeros of polynomials of degree larger that 4. We may, nevertheless, infer their general form as follows. We first notice that  ${\cal{P}}_{\ell}(r)$ may be written in the form,
\begin{equation}\label{aw14}
{\cal{P}}_{\ell}(r)= \frac{\prod_{k=1}^{\ell+2} (r-r_k)}{\prod_{k=1}^{\ell+2} (-r_k)}
\end{equation}
where $r_k$ are the zeros of ${\cal{P}}_{\ell}(r)$, which, as indicated are simple. Therefore, since $y^2-2My+Q^2 = (y-r_i)(y-r_o)$, we should have,
\begin{eqnarray}
\label{aw15}
  \int_0^r \frac{ y^4}{(y^2-2My+Q^2)\left({\cal{P}}_{\ell}(y)\right)^2} dy &=& A \ln(r_i-r)+B\ln(r_o-r) + C
\nonumber \\ & & +\sum_{k=1}^{\ell+2} a_i \ln(r-r_k)
    + \sum_{k=1}^{\ell+2}{\frac{b_i}{(r-r_k)}}
\end{eqnarray}
where $A$, $B$, $C$, $a_k$ and $b_k$ are constants that depend on $M$, $Q$, and  $r_k$. The last term in (\ref{aw15}) may be written in the form,
\begin{equation}\label{aw16}
\sum_{k=1}^{\ell+2}{\frac{b_i}{(r-r_k)}}= \frac{U_{\ell}(r)}{{\cal{P}}_{\ell}(r)}
\end{equation}
where $U_{\ell}(r)$ is a polynomial of order $\ell-1$, or lower. Replacing in (\ref{aw11}),
\begin{eqnarray}
\label{aw17}
  \psi_b(r) &=& \frac{1}{r} {\cal{P}}_{\ell}(r)\left(A \ln(r_i-r)+B\ln(r_o-r) + C\right)  + \frac{1}{r} {U}_{\ell}(r) \nonumber \\
   & & +\frac{1}{r} {\cal{P}}_{\ell}(r)\sum_{k=1}^{\ell+2} a_i \ln(r-r_k)
\end{eqnarray}
But we notice that $\psi_b(r)$ is a solution of (\ref{aw01}), which can be singular only at the regular
singular points $r=0$, $r=r_i$, and $r=r_o$, and that the zeros $r_k$ do not coincide with these points.
Therefore, we must have $a_k=0$ for all $k$, and the last term in (\ref{aw17}) vanishes identically. This
result implies that, for any $\ell$, we may construct algebraically the solution (\ref{aw11}) as follows.
 We first compute the coefficients of ${\cal{P}}_{\ell}(r)$ as indicated above, and then replace in (\ref{aw17})
leaving $A$, $B$, $C$ and the coefficients of $U_{\ell}$ arbitrary. Next we replace in (\ref{aw01}) and impose
the condition that $\psi$ is a solution of that equation, and that $\psi \sim r^4$ near $r=0$. It can be
 checked that this procedure determines all the coefficients, up to an arbitrary multiplicative constant,
a simple example being (\ref{l2zvm}) for $\ell=2$. Since by construction these solutions satisfy the
appropriate boundary condition at $r=0$, the construction of the corresponding {\em scalar} zero modes,
and the associated intertwining potential, is now a simple algebraic procedure. The resulting expressions
are, unfortunately, very long and rather difficult to analyze in detail. In particular, we have not been
able to show explicitly that for $0 < r_c < r_i$ the scalar zero modes that satisfy the required boundary
 condition at $r=0$ are non vanishing everywhere in the interval $0 < r < r_i$, as required for the
regularity of the intertwining potential. We remark, nevertheless, that this appears to be the case
in all the particular solutions analyzed numerically after assigning definite numerical values for
the parameters, as in the example described in Section III - C.

\end{document}